\documentclass[referee]{an}
\usepackage{graphicx}
\usepackage{times}
\usepackage[dvips]{epsfig}

\begin{document}

% The following seven commands are intended for editorial usage and
% should be ignored by the author(s).
\Pagespan{1}{}% Document's page range. 
% If second parameter is left empty, the last page is computed
% automatically.
\Yearpublication{2019}%
\Yearsubmission{2019}%
\Month{1}%   
\Volume{999}%  
\Issue{92}% 
% \DOI{This.is/not.aDOI}% 

%%%%%%%%%%%%%%%%%%%%%%%%%%%%%%%%%%%%%%%%%%%%%%%%%%%%%%%%%%%%%%%%%%%%%%%%%%%%%%%%%%%%%%%%%%%%%%%%%%
\title{The effects of a compressive velocity pulse on a collapsing turbulent clump}

\author{G. Arreaga-Garc\'{\i}a\inst{1}\fnmsep\thanks{Corresponding author:
\email{guillermo.arreaga@unison.mx}}
% Example for footnote, note the usage of the \texttt{fnmsep} command
% as separator between institute number and footnote mark}
}
\titlerunning{compressive wave on a turbulent clump}
\authorrunning{Arreaga-Garcia}
\institute{Departamento de Investigaci\'on en F\'{\i}sica de la
Universidad de Sonora, M\'exico, Apdo. Postal 14740, Hermosillo, 83000 Sonora, M\'exico.}
%%%%%%%%%%%%%%%%%%%%%%%%%%%%%%%%%%%%%%%%%%%%%%%%%%%%%%%%%%%%%%%%%%%%%%%%%%%%%%%%%%%%%%%%%%%%%%%%
\received{XXXX}
\accepted{XXXX}
\publonline{XXXX}
%%%%%%%%%%%%%%%%%%%%%%%%%%%%%%%%%%%%%%%%%%%%%%%%%%%%%%%%%%%%%%%%%%%%%%%%%%%%%%%%%%%%%%%%%%%%%%%%
\keywords{--stars: formation; 
          --physical processes: gravitational collapse, hydrodynamics; 
          --methods: numerical;}
%%%%%%%%%%%%%%%%%%%%%%%%%%%%%%%%%%%%%%%%%%%%%%%%%%%%%%%%%%%%%%%%%%%%%%%%%%%%%%%%%%%%%%%%%%%%%%%%
\abstract{High-resolution hydrodynamical simulations are presented to 
follow the gravitational collapse of a uniform turbulent 
clump, upon which a purely radial compressive velocity pulse 
is activated in the midst of the evolution of the clump, when its turbulent state has 
been fully developed. The shape of the velocity pulse is determined basically 
by two free parameters: the velocity $V_0$ and the initial 
radial position $r_0$. In the present paper, models are considered 
in which the velocity $V_0$ takes the values 2, 5, 10, 20, and 50 times 
the speed of sound of the clump $c_0$, while $r_0$ is fixed for all the models. The collapse 
of the model with $2 \, c_0$ goes 
faster as a consequence of the velocity pulse, while the cluster formed in the central 
region of the isolated clump mainly stays the same. In the models with greater velocity $V_0$,
the evolution of the isolated clump is significantly changed, so that a dense 
shell of gas forms around $r_0$ and moves radially inward. The radial 
profile of the density and velocity as well as the mass contained in the dense 
shell of gas are calculated, and it is found that (i) the higher the 
velocity $V_0$, the less mass is contained in the shell; (ii) there is a critical velocity of 
the pulse, around $10 \, c_0$, such that for shock models with a lower velocity, 
there will be a well defined dense central region in the shocked clump surrounded by the 
shell.}

%%%%%%%%%%%%%%%%%%%%%%%%%%%%%%%%%%%%%%%%%%%%%%%%%%%%%%%%%%%%%%%%%%%%%%%%%%%%%%%%%%%%%%%%%%%%%%%
\maketitle
%%%%%%%%%%%%%%%%%%%%%%%%%%%%%%%%%%%%%%%%%%%%%%%%%%%%%%%%%%%%%%%%%%%%%%%%%%%%%%%%%%%%%%%%%%%%%%%
\section{Introduction}
\label{sec:intro}

The propagation of shock waves arises in many astrophysical 
systems. In particular, there is ample observational evidence on the interaction 
of a shock with a gas structure, see for instance \cite{hwang}, who reported 
{\it Chandra} images of a shocked cloud being torn apart by shear instabilities 
in the Puppis A supernova remnant.

Further observational evidence of the shock--gas interaction has been 
provided by \cite{nutter}, who 
presented submillimetre measurements to explain the difference in the star formation 
activity of the clumps L1688 and L1689 of the $\rho$ Ophiuchi molecular cloud 
complex. Indeed, these authors proposed that both clumps are being affected by members of 
the Upper Scorpius OB association, so that there is triggered star formation at different 
rates because the distance from the clumps to the most massive and luminous nearby 
component of the Upper Scorpius OB association is different.

In addition, interferometric observations have revealed that the cores L1689B 
and L694-2 seem to collapse faster than their spontaneous collapse would 
indicate, as their infall velocities were observed to be faster 
than expected, see \cite{lee}. Thus, it was 
theoretically suggested by \cite{seo} that these cores may be strongly 
influenced by an external factor, for example, turbulence, or an 
external pressure.  

With regard to the theoretical aspect, papers on computational 
simulations of the interaction of a shock and gas 
clouds started to appear many decades ago, see for 
instance \cite{stone}. These authors considered a 
10 Mach shock that impacted on a cloud, so that its three-dimensional 
evolution led to its total disruption. More recently, other 
authors have also found that gas clouds are destroyed by an 
incident shock wave, among others, see \cite{naka06}, 
\cite{pittard09, pittard10, pittard16}. 

The previous studies of \cite{stone,naka06,pittard09, pittard10, pittard16} are concerned
with the destruction of interstellar clouds by shock waves. A more interesting scenario 
for star formation is based on the idea that 
a bound gas structure can be compressed by a shock to the point that its collapse can 
be triggered. In fact, \cite{boss95} investigated this possibility, so that a radially 
inward velocity perturbation induced the collapse of a rotating centrally condensed 
three-dimensional core. In addition, \cite{vanhala} demonstrated that the interaction of 
a planar shock wave with a centrally concentrated three-dimensional cloud is capable 
of triggering the gravitational collapse of the cloud, when radiative cooling in their 
smoothed particle hydrodynamics (SPH) simulations was taken into account. 

Subsequently, using simulations limited to 
one radial dimension, \cite{gomez2007} investigated whether a velocity pulse can trigger 
the formation of prestellar core when it strikes a 
gas cloud. These authors proposed a mathematical function such that the 
shape of the velocity pulse is mainly determined by 
the following parameters: the velocity $V_0$, the radii $r_0$ and $r_1$, and the increments on 
these radii $\delta r_0$ and $\delta r_1$, all of which are described in 
Section~\ref{subsec:velpulse} of the present paper. These authors acknowledge that their setup 
was somewhat unphysical since it restricts the nature of the compressive wave to ''one-dimensional'' spherical shells.   

In this paper, three-dimensional high-resolution hydrodynamical 
simulations of the interaction of the velocity pulse proposed 
by \cite{gomez2007} with a turbulent clump are presented. Models 
are considered in which the velocity $V_0$ takes the 
values 2, 5, 10, 20 and 50 times the speed of sound of the 
clump, while the other parameters are fixed.

It should be emphasized that consequently with the previous hypothesis, the initial energies of the 
isolated turbulent clump of this paper are chosen to make it to collapse spontaneously, even without 
the compressive velocity pulse. The papers quoted previously of other authors 
like \cite{stone,naka06,pittard09, pittard10, pittard16}, investigated the effect of shock waves on 
clumps and clouds that would otherwise not collapse.

It should be emphasized that \cite{henne} investigated numerically the effects of a steady increase 
in the external pressure on the collapse of 
a prestellar core. For slow to moderate compression rates, subsonic infall velocities were observed to develop in the 
outer parts of the core. These authors also found that ''a compression wave is driven into the core, thereby 
triggering collapse from the outside in''. This observation, together with the model of \cite{gomez2007}, have been 
the physical motivation of this paper, so that we have an incident three-dimensional 
spherical velocity pulse to simulate the effect of a larger, already collapsing 
cloud on an embedded turbulent clump.   
 
In this paper, the kinetic energy of the clump is provided by means of a
velocity field assigned according to a decaying, curl-free turbulent spectrum. At the 
end of the simulation run, a lot of fragmentation is seen to occur in the central region of the 
clump, as was previously observed in other simulations, see for instance \cite{offner2009} 
and \cite{offner2010}. Later, when the first stage of evolution of the clump has passed and 
the turbulence has reached a fully developed state, a radially inward velocity pulse is 
activated, so that its effects on the subsequent evolution of the clump can be studied. 

Consequently, the present paper 
is similar to that of \cite{offner2009} and \cite{offner2010} with the addition of a spherical compressive velocity 
pulse, executed with the fully-parallelized 
publicly available code Gadget2, which is based on the Smoothed Particle Hydrodynamics (SPH) 
technique, see \cite{gadget2} and \cite{serial}. 

In addition, when gravity has produced a 
substantial contraction of the clump, the gas begins to heat, so that its increase of temperature 
is taken into account by means of a barotropic equation of state proposed by \cite{boss2000}.      
It should be mentioned 
that the equation of state is a very important factor on triggered star 
formation in molecular clouds, as the post-shock densities depend on the square of the 
Mach number of the shock wave, and therefore higher compression factors can be reached 
in isothermal simulations than in the adiabatic case. \cite{fosterbossa,fosterbossb} concluded that for 
high-velocity shocks (greater than 100 km/s) the post-shock material is heated and becomes adiabatic. In these 
cases, the cloud can be destroyed. For low-velocity shocks, the post-shock material remains isothermal and 
high compression factors can also be achieved: self-gravity then leads to collapse. It was soon realized that 
the velocity of the shock together with the equation of state can make the difference between the two 
scenarios mentioned above, that is, triggered collapse or destruction of the 
gas structure.       

The outline of the paper is as follows. In Section \ref{sec:met}, the physical state of the clump, including 
its turbulent velocity spectrum, its initial energy conditions; the implementation of the velocity pulse and some computational 
details of the evolution are described. In Section \ref{sec:results}, the main results of the simulations are presented 
by means of iso-density and velocity plots. In Sections \ref{sec:den2} and \ref{sec:themass}, a 
physical characterization of the simulation output is reported by means of plots of the density and 
velocity radial profile. In Section \ref{sec:dist}, a comparison between models with an equal distance from the shell of dense 
gas to the clump centre is presented. Finally, in Section 
\ref{sec:dis} and \ref{sec:conclu}, the main results and conclusions are summarized.  
%%%%%%%%%%%%%%%%%%%%%%%%%%%%%%%%%%%%%%%%%%%%%%%%%%%%%%%%%%%%%%%%%%%%%%%%%%%%
%%%%%%%%%%%%%%%%%%%%%%%%%%%%%%%%%%%%%%%%%%%%%%%%%%%%%%
\section{The physical system and the computational method}
\label{sec:met}

%%%%%%%%%%%%%%%%%%%%%%%%%%%%%%%%%%%%%%%%%%%%%%%%%%%%%
%%%%%%%%%%%%%%%%%%%%%%%%%%%%%%%%%%%%%%%%%%%%%%%%%%%%%%
\subsection{The clump}
\label{subsec:clump}

The gas structure considered in this paper has a radius of 
$R_0$=2.0  $\times 10^{18} \,$ cm
$\equiv$ 0.65 pc and a mass of $M_0$=185 $M_{\odot}$. Thus, the 
average density and the corresponding free fall time of this gas structure are 
$\rho_0$=1.08 $\times 10^{-20}$ g cm$^{-3}$ and 
$t_{ff} \approx $2.01 $\times 10^{13} \,$ s $\equiv$ 0.63 Myr, 
respectively. The values of $R_0$ and $M_0$ are very similar to those used 
by \cite{offner2009} and \cite{offner2010}. 

According to the naming convention introduced 
by \cite{bergin}, which is based on the mass and size 
of the gas structures, the one considered in the present paper is usually referred to 
as a gas clump. The physical state of the clump and the initial conditions 
implemented are next explained, in Sections~\ref{subsec:velpulse}--\ref{subs:code}. 
 
%%%%%%%%%%%%%%%%%%%%%%%%%%%%%%%%%%%%%%%%%%%%%%%%%%%%%%%%%%%%%%%
\subsection{The velocity pulse}
\label{subsec:velpulse}

The mathematical function for describing the velocity pulse has been 
taken from Eq. 2 of \cite{gomez2007}, which is given by\footnote{There is presumably a mistake (typo) in 
Eq. 2 of \cite{gomez2007}, so that in this paper we changed the sign.} 
 
\begin{equation}
\begin{array}{l}

v(r) \, \approx 

\left\{

\begin{array}{l}

0 \; \mbox{for} \; r < r_0-\delta r_0 \\

V_0 \;  \sin\left[ \frac{\pi}{2} \left( \frac{r-r_0}{\delta r_0 }\right) \right] 
\; \mbox{for}  \; r_0-\delta r_0 < r < r_0 + \delta r_0  \vspace{0.25 cm} \\

V_0 \; \sin\left[ \frac{\pi}{2} \left( \frac{r_1-r}{\delta r_1 }\right) \right]  
\; \mbox{for} \;  r > r_0+\delta r_0\\
\end{array}

\right.
\end{array}
\label{velpulse}
\end{equation}
\noindent where $V_0$ determines the amplitude of the wave, for which it is the most 
important parameter, so that it will be varied later to define the models, see Table \ref{tab:mod} 
of Section \ref{sec:results}. Other parameters are $r_0$, $r_1$ which determine the 
initial radii where the pulse will be activated, and they are fixed as $r_0=0.8$, 
and $r_1=\frac{1 + \left( r_0+ \delta r_0 \right)}{2}=0.95$, where the width of the wave 
is controlled by  the parameters $\delta r_0=0.1$ and 
$\delta r_1=\frac{1 - \left( r_0+ \delta r_0 \right)}{2}=0.05$. It should be emphasized that 
the radii $r_0$, $r_1$, $\delta r_0$ and $\delta r_1$ are all given in terms of the initial 
radius of the clump, $R_0$. 

The function $v(r)$ can be taken entirely as the radially inward component of the velocity 
field, so that this paper can be considered as (i) another implementation of the idea previously 
considered by \cite{boss95} (using another mathematical function) and (ii) a generalization to 
three-dimensions of the one-dimensional simulations calculated by \cite{gomez2007}. 

Let the coordinates of a particle be given by $(r,\theta,\phi)$, where 
$r$ is the radial distance, $\theta$ is the polar angle and $\phi$ is the 
azimuthal angle. Then, the relation between the velocity components 
in spherical and Cartesian coordinates is

\begin{equation}
\begin{array}{l}
v_r = \sin(\theta) \, \cos(\phi) \, v_x + \sin(\theta) \, \sin(\phi) \, v_y + \cos(\theta)\, v_z \\   
v_{\theta} = \cos(\theta) \, \cos(\phi) \, v_x + \cos(\theta) \, \sin(\phi) \, v_y - \cos(\theta)\, v_z \\ 
v_{\phi}= - \sin(\phi) \, v_x + \cos(\phi) \, v_y.
\end{array}
\label{eqsimulvel}
\end{equation}
\noindent so that it can be assumed that the left hand side of 
Eq.~\ref{eqsimulvel} is given by $v_r= v(r)$, as shown 
in Eq.~\ref{velpulse}, while the polar and azimuthal components are $v_{\theta} \approx 0 $ 
and $v_{\theta} \approx 0$, respectively.
 
\begin{figure}
\begin{center}
\begin{tabular}{cc}
\hspace{-1 cm } \includegraphics[width=4.0 in]{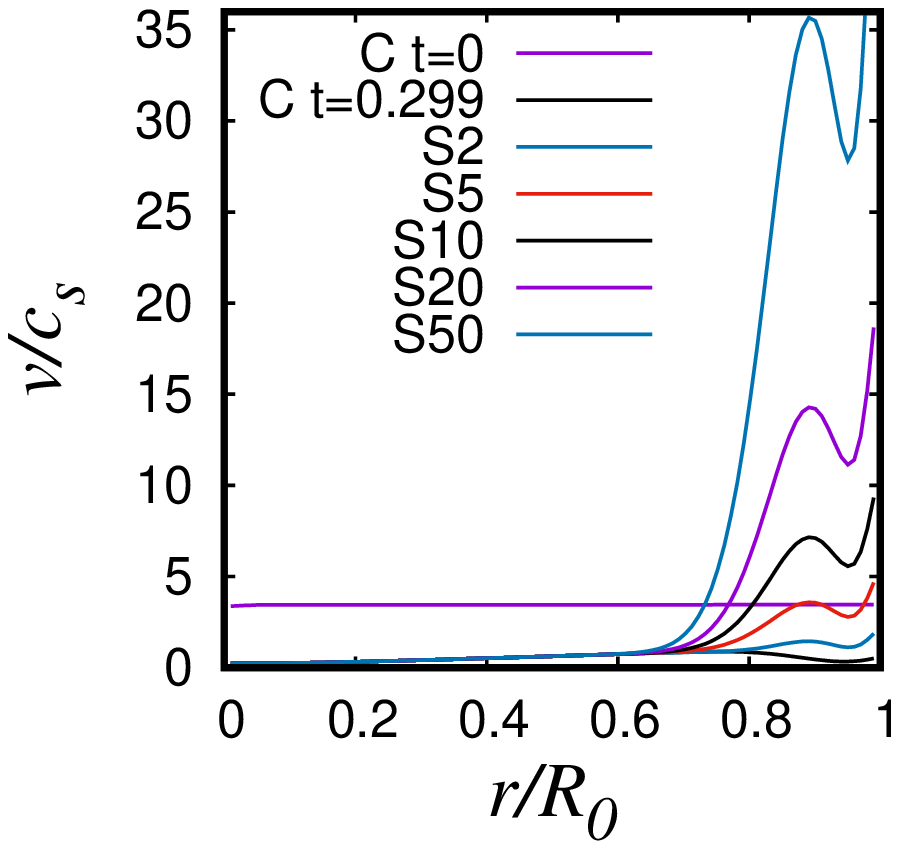} & \hspace{-3 cm} \includegraphics[width=4.0 in]{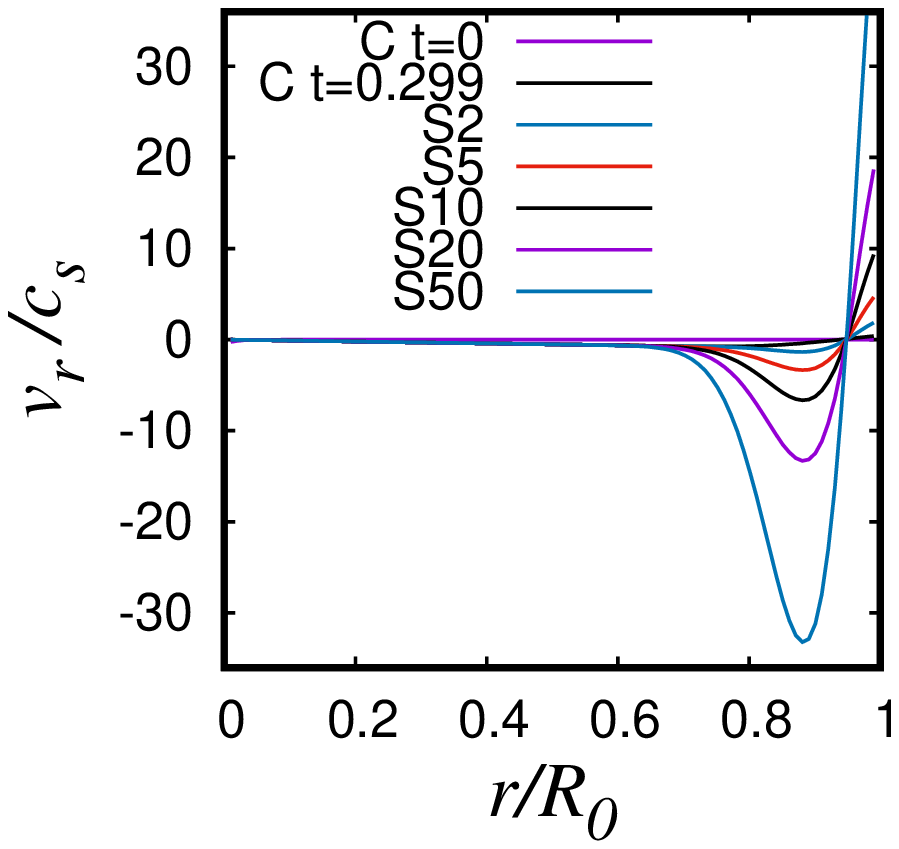}
\end{tabular}
\caption{\label{velCurf2S} For the first snapshot of each model, the measured velocity profile 
against the clump radius for Models C (two curves at times $t/t_{ff}$=0 and $t/t_{ff}$=0.299) and 
S2--S50, is shown by plots of the (left) velocity magnitude and (right) radial 
component of the velocity. Velocities are normalized with the initial 
speed of sound of the clump.}
\end{center}
\end{figure}

Thus, the Gauss--Jordan elimination method described in \cite{numrecip} can be used to solve 
these equations simultaneously for each simulation particle in order to get the 
Cartesian components $v_x, v_y$, and $v_z$, which are needed for 
the evolution code, see Section~\ref{subs:code}. The velocities of the gas particles 
actually given to the Gadget2 code at the initial time are shown in 
Fig.~\ref{velCurf2S}\footnote{For details about the radial partition of these plots, see 
Section~\ref{sec:den2}.}. 

Let us consider now Fig.~\ref{velCurf2S}. The curves with label C t=0 describe 
the radial velocity profile of the clump at the initial evolution time, so that the 
homogeneous distribution of matter is evident. At the time $t/t_{ff}=$0.299, the 
matter distribution of the clump has obviously changed as the curves with label C t=0.299 
indicate. It should be noted that the outermost particles of the clump expand outwards 
spontaneously, as can be seen in the right panel of 
Fig.~\ref{velCurf2S}. This happens because the turbulent clump is not in hydrodynamic equilibrium and 
there is no external pressure acting upon it. However, this expansion does not take place only 
radially, as the outermost particles may have a non-zero tangential component of the velocity in addition 
to the radial component already seen in this panel.            

However, only those particles within $r_0-\delta r_0 < r < r_0 +\delta r_0$ (an interval of $\approx 0.8 < r < 0.95$) 
are part of the velocity pulse, so that they have been assigned a radially 
inward velocity. Because of this, a discontinuity 
in the tangent component of the velocity has been imposed, as can be seen in the small fall in the magnitude 
of the velocity (shown in the left panel of Fig.~\ref{velCurf2S}) exactly at the radius 
($\approx$ 0.95) where the radial component of the velocity (shown in the right panel of Fig.~\ref{velCurf2S}) 
changes their sign from negative to positive values.
%%%%%%%%%%%%%%%%%%%%%%%%%%%%%%%%%%%%%%%%%%%%%%%%%%%%%%%%%%%%%%%%%%%%%%%%%%%%%%%%%%%%%%%%%%%%%%%%%%%%%%%%%%         
\subsection{The turbulent velocity}
\label{subsec:velfin}

In order to generate the turbulent velocity spectrum, a 
mesh with a side length equal to the clump radius $R_0$ is used, so that the size of 
each grid element of this mesh is $\delta = R_0/N_g$ and the mesh partition is
determined by $N_g=128$. In Fourier space, the partition is given by $\delta K= 1/R_0$, so that
each wavenumber $\vec{K}$ has the components
$K_x=i_{x}  \delta K$, $K_y=i_{y}  \delta K$ and $K_z=i_{z}  \delta K$
where the indices $i_x,i_y,i_z$  take integer values in
the range $[-N_g/2,N_g/2]$ to cover all the mesh.

The initial power of the velocity field must be given by

\begin{equation}
P(\vec{K})=\left< \left| \vec{v}(\vec{K}) \right|^2 \right> \approx \left| \vec{K}\right|^{-n}
\label{power}
\end{equation}
\noindent where the spectral index $n$ is a constant, $n=-1$. Thus, 
following \cite{dobbs}, \cite{RMAA2017} and \cite{miAAS18}, the components of the 
particle velocity, in the case of a diverge-free turbulent 
spectrum, are given by

\begin{equation}
\vec{v}_{\rm CF}(\vec{r})  \approx \Sigma_{i_x,i_y,i_z} \left| \vec{K}\right|^{\frac{-n-2}{2}} \;
\vec{K} \sin \left( \vec{K}\cdot \vec{r} + \Phi_K \right)
\label{velturb}
\end{equation}
\noindent where $\Phi_{K_x},\Phi_{K_x}$ and $\Phi_{K_z}$ are random phases of the wave 
and $\vec{r}$ is the particle position in real space.     

%%%%%%%%%%%%%%%%%%%%%%%%%%%%%%%%%%%%%%%%%
\subsection{Initial energies}
\label{subs:energies}

In the standard SPH formulation, the thermal, kinetic and gravitational energies 
of the set of gas particles are calculated by 

\begin{equation}
\begin{array}{l}
E_{\rm ther}=\frac{3}{2}\sum_{i} \, m_{i}\frac{P_{i}}{\rho _{i}}\\
E_{\rm kin}=\frac{1}{2}\sum_{i} \, m_{i} v_i^{2},\\
E_{\rm grav}=\frac{1}{2}\sum_{i} \, m_{i}\Phi_{i}
\label{energiespart}
\end{array}
\end{equation}
\noindent where $P_i$ is the pressure and $\Phi _{i}$ is the gravitational 
potential at the location of particle $i$, with velocity $v_i$ and mass $m_i$, to be 
defined in Section~\ref{subs:code}. It should be kept in mind that all the SPH 
particles of a simulation must be used in the summation of Eq.~\ref{energiespart}.

Now, let $\alpha$ be defined as the ratio of the thermal energy to the gravitational
energy and let $\beta$ be the ratio of the kinetic energy
to the gravitational energy, so that

\begin{equation}
\alpha \equiv \frac{E_{ther}}{\left|E_{grav}\right|}
\label{defalpha}
\end{equation}
\noindent and

\begin{equation}
\beta \equiv \frac{E_{kin}}{\left|E_{grav}\right|}.
\label{defbeta}
\end{equation}

In this paper, the value of the speed of sound $c_0$ is fixed at 34247.56 cm/s, so that

\begin{equation}
\alpha \equiv 0.24
\label{valpha}
\end{equation}
\noindent for all the shock models and for the isolated clump model as well. 

Another very useful quantity to characterize the physical 
state of a gas structure is the so-called virial parameter, which is defined observationally by 

\begin{equation}
\beta_{\rm vir} \equiv \frac{5 \, \sigma_{1D}^2 \, R}{G \, M }
\label{defbetavir}
\end{equation}
\noindent where $G$ is Newton's gravitational constant, $M$ and $R$ are the mass and radius 
of a general gas structure, and $\sigma_{1D}$ is the intrinsic one-dimensional velocity dispersion of 
the hydrogen molecule of mean mass. Assuming isotropic motions, a three-dimensional velocity dispersion 
can be simply estimated by $\sigma_{3D}= \sqrt{3} \, \sigma_{1D}$. It should be noted that a gas structure in virial
equilibrium would have  

\begin{equation}
\beta_{\rm vir}=1  
\label{vbetavir}
\end{equation}     
\noindent so that, using the mass and radius of the clump described in Section \ref{subsec:clump} 
in Eq.\ref{defbetavir}, together 
with Eq.\ref{vbetavir}, a value of $\sigma_{1D}=1.44$ km/s is obtained. 

The magnitude of the velocity field described in 
Eq.\ref{velturb} of Section \ref{subsec:velfin}, has been calibrated so that the three-dimensional 
velocity dispersion of the simulation particles, $\sigma_{3D}^s$ has been 
calculated to be equal to $\sigma_{1D}$, so that $\sigma_{3D}^s=1.44$ km/s. In this case, the approximate virial 
parameter of the turbulent clump described in Section \ref{subsec:clump} and \ref{subsec:velfin}, would initially
take a value a little bit below 1.

In any case, when the velocity pulse is activated in the turbulent clump, there will be a value 
of $\beta_{\rm vir}$ and $\beta$ for each model. Only $\beta$ is presented in 
Table~\ref{tab:mod} with the purpose of characterizing further the dynamical state of 
the corresponding clump and shock models as well.

\cite{xing} reported the values of the virial parameter  
observed in many gas cloud cores, which in general were found to be smaller than or around the value 1. In 
addition, \cite{bertoldi} found 
that the observed virial parameter depends on the total mass of the gas structure. \cite{kauffmann} 
discussed how the value of the 
virial parameter determines the later evolution of a self-gravitating,  
non-magnetized gas structure: a virial parameter above 2 indicates that the gas structure
is unbound and may expand, while one below 2 suggests that the gas structure is bound and 
may collapse.

Theorists have often used another expression for the virial parameter, for instance, $\beta_{\rm vir}= 2 \, a \, \beta$, 
where $\beta$ is the dimensionless ratio defined in Eq.~\ref{defbeta} and $a$ is a numerical 
factor which is empirically included to take into account modifications of non-homogeneous 
and non-spherical density distributions.         

It must be recalled that the important issue for the present paper is to have initially a globally 
collapsing clump, as was mentioned in Section \ref{sec:intro}. In addition, it must be mentioned that \cite{peretto} 
observed the clump NGC 2264-C and found it to be in a global state of collapse, far from hydrostatic 
equilibrium, as its virial parameter is around 0.2, which indicates that the clump is very unstable 
gravitationally.
%%%%%%%%%%%%%%%%%%%%%%%%%%%%%%%%%%%%%%%%%%%%%%%%%%%%%%%%%%%%%%%%%%%%%%%%%%%%
%%%%%%%%%%%%%%%%%%%%%%%%%%%%%%%%%%%%%%%%%%%%%%%%%%%%%%%%%%%%%%%%%%%%%%%%%%%%%%%%%%%%%%%%%%%%%%%%%%%%%%
\subsection{Evolution code, resolution and equation of state}
\label{subs:code}

The temporal evolution of all the models of the present paper has been solved using 
the particle-based computer program Gadget2, see~\cite{gadget2} and 
also \cite{serial}. Gadget2 is based on the
tree-PM method for computing the gravitational forces and on the
standard SPH  method for solving
the Euler equations of hydrodynamics. Gadget2 implements a
Monaghan--Balsara form for the artificial viscosity;
see~\cite{mona1983} and \cite{balsara1995}. The strength of the
viscosity is regulated by the parameter $\alpha_{\nu} = 0.75$ and
$\beta_{\nu}=\frac{1}{2}\, \times \alpha_v$; where $\alpha_{\nu}$ and $\beta_{\nu}$ are coefficients of linear 
and quadratic viscosity; see Eqs~11 and 14
in~\cite{gadget2}. In the simulations presented in this paper, the 
Courant factor is fixed at $0.1$.

Following \cite{truelove} and \cite{bateburkert97}, the smallest mass of a particle 
that the SPH calculation must resolve in order to be reliable is given 
by $\frac{m_p}{m_r}<1$, where $m_p$ is the mass of the simulation particle 
and $m_r \approx M_J / (2 N_{\rm neigh})$, so that
$N_{\rm neigh}$ is the number of neighbouring particles included in the
SPH kernel and $M_J$ is the Jeans mass, which can be expressed by 

\begin{equation}
M_J \equiv \frac{4}{3}\pi \; \rho \left(\frac{ \lambda_J}{2}
\right)^3 = \frac{ \pi^\frac{5}{2} }{6} \frac{c^3}{ \sqrt{G^3 \,
\rho} } 
\label{mjeans}
\end{equation}
\noindent where $\lambda_J$ is the Jeans wavelength, $c$ is the instantaneous speed of 
sound, $\rho$ is the local density, and $G$ is Newton's gravitation constant. In addition, 
the values of the density and speed of sound must be updated according to the 
following equation of state:

\begin{equation}
p= c^2 \, \rho \left[ 1 + \left(
\frac{\rho}{\rho_{crit}}\right)^{\gamma -1 } \, \right] ,
\label{beos}
\end{equation}
\noindent which was proposed by \cite{boss2000}, where $\gamma\,
\equiv 5/3$ and for the critical density the 
value $\rho_{crit}=5.0 \times 10^{-14} \, $ g $\,$ cm$^{-3}$ is assumed. 

Let us say something about 
this equation of state. The ideal equation of state is a good approximation to the thermodynamics of the observed
star forming regions, which basically consist of 
molecular hydrogen cores at $10\,$ K with an average density of $1 \times 10^{-20} \, $ g $\,$ cm$^{-3}$.
Furthermore, once gravity has produced a substantial core contraction, the gas 
begins to heat. Therefore, the most important free parameter is then
the critical density $\rho_{crit}$, which determines the change of thermodynamic
regime from isothermal to adiabatic. It should be emphasized that this approach is introduced with 
the purpose of avoiding the consideration of a radiative transfer problem coupled with an energy equation in the 
hydrodynamics equations. Indeed, \cite{white} studied the 
collapse of cores including radiative transfer in the flux-limited diffusion approximation. These authors observed 
important differences in the dependence of temperature on density. However, \cite{arreaga2008} compared the results 
of simulations of the collapse of the uniform density core with those of \cite{white} at several values 
of $\rho_{crit}$ and found that the barotropic equation of state behaves quite well within the range of $\rho_{crit}$ 
between $10^{-14}-10^{-11}$ g $\,$ cm$^{-3}$. Based on these results, in this paper the parameter 
$\rho_{crit}$ has been fixed at the value mentioned above.

Assume that a typical peak density of 
$\rho=1.0 \times 10^{-11} \, $ g $\,$ cm$^{-3}$ can be reached in the late evolution of 
the shock models and for the turbulent clump considered in the present paper 
as well, so that the total number of particles in a simulation is $N_p=$15,417,343. Thus, 
the particle mass is given by $m_p=1.19 \times \, 10^{-5} \,  M_{\odot}$ and the Jeans mass 
is $M_j \approx 0.001 \, M_{\odot}$, so that the
minimum mass is $m_r \approx 1.34 \times 10^{-5}\, M_{\odot}$, since $N_{\rm neigh}=40$. Therefore, the
ratio of masses is given by $m_p/m_r=$0.89 and then the desired
resolution is achieved in our simulations.

Thus, for the stated average density $\rho_0$ of the clump, in this calculation it will be initially 
isothermal and only when the peak density has increased by six orders of magnitude, it will become adiabatic.   
%%%%%%%%%%%%%%%%%%%%%%%%%%%%%%%%%%%%%%%%%%%%%%%%%%%%%%%%%%%%%%%%%%%%%%%%%%%%%%%%%%%%%%%%%%%%%%%%%%%%%%%%%%%%%%%%%%%%%%
\section{Results}
\label{sec:results}

\subsection{Visualization}
\label{subs:vis}

The main outcome of each model is illustrated by
means of a iso-density plot, in which a slice of the simulation 
particles is used from the last snapshot 
available. The width of the slice is determined so that it contains around 
10,000 gas particles; the characteristic width of the slice in terms of the clump 
radius $R_0$ is of the order of $3\, \times 10^{-4}$. In addition, the slice is parallel to 
the $x-y$ plane and its height along the 
the $z$-axis is determined by the $z$-coordinate of the highest density particle. A bar located 
at the bottom of each iso-density plot shows the range of values for the base 
10 logarithm of the ratio of the peak iso density to the average 
density of the initial clump, that is, $\log_{10} \left( \rho_{\rm max}(t)/\rho_0 \right)$. 
It must be emphasized that this logarithm scale of density is used to (i) set a gray 
scale, so that the different density regions can be distinguished in the iso-density plots 
to show in Sections \ref{subs:modelc}--\ref{subs:modelS50}; (ii) set a variation scale of density, so that a density width can be defined 
in order to determine the mass contained in a shell of dense gas, see Section \ref{sec:themass}; (iii) follow the position of 
the shell of dense gas with respect to the clump centre for all the evolution time, see Section \ref{sec:dist}.       

It should  be explained that the vertical and horizontal axes of all
the iso-density plots indicate the length in terms of the radius
$R_0$ of the clump. Hence, the Cartesian $x$- and $y$-axes
vary initially from -1 to 1. However, in order to
facilitate the visualization of the last configuration obtained, the same length scale per 
side is not used in all the plots. In addition, for each iso-density plot, a velocity 
plot is also presented, which is constructed using 
the same slice of simulation particles, so that an arrow is located at the site of each particle, with 
its length being proportional to the total magnitude of the velocity.
 
The global evolution of the models will be described in Sections~\ref{subs:modelc} 
and \ref{subs:overe}. As desired, all the models collapse, so that the peak density reached 
at every point in time of the evolution (where a snapshot of each model is available) is determined and shown 
in Fig.~\ref{fDenMax}. Later, in order to characterize 
further the shock models, the behaviour of the radial profile of the density and velocity of the particles  
will be calculated and described in Section~\ref{sec:den2}.   

%%%%%%%%%%%%%%%%%%%%%%%%%%%%%%%%%%%%%%%%%%%%%%%%%%%%%%%%%%%%%%%%%%%%%%%%%%%%%%%%%%%%%%%%%%%%%%%%%%%%%%%%%%%%%
\subsection{The clump model}
\label{subs:modelc}

It is well known that a turbulent velocity spectrum, such as the one implemented 
here, induces a lot of collisions between the particles, so that the 
density of some particles will be increased, in many places of the 
clump simultaneously. The initial compressive velocity pulse will steepen immediately 
into an inward-moving shock wave. Since the equation of state is initially isothermal, the compression factor
will depend on the Mach number squared and thus the higher velocity shock waves will produce the 
highest post-shock densities, as is predicted by the so-called 
Rankine-Hugoniot jump conditions, see \cite{jumpc}. 

Such behaviour is manifested 
in the oscillations of the density curves 
shown in Fig.~\ref{fDenMax}, so that the first stage of the evolution 
of the turbulent clump can be defined in this way.      

The second stage can be identified at 
times beyond $0.1 \, t_{ff}$, in which the clump becomes more 
homogeneous, so that the peak density curve stretches almost horizontally, up to 
a time of around $0.8 \, t_{ff}$; the third stage can be seen at times $t$ 
within $0.8 \, t_{ff} < t < 1.0 \, t_{ff}$, at which a true collapse of the 
clump takes place, in a collapse time a little bit smaller than $t_{ff}$.

The evolution of the isolated turbulent clump has been followed up to the point when a lot 
of fragmentation is produced in its central region, so that a cluster of small 
over-densities can be seen in Fig.~\ref{fCurf2final}. Similar results have been obtained by many authors, so 
that there are papers studying the statistical properties of the resulting fragments, see for 
instance \cite{bate2003, bate2012}.   
 
%%%%%%%%%%%%%%%%%%%%%%%%%%%%%%%%%%%%%%%%%%%%%%%%%%%%%%%%%%%%%%%%%%%%%%%%%%%%%%%%%%%%%%%%%%%%%%%%%%%%%%%%%%
\subsection{The overall evolution of the shocked clump}
\label{subs:overe}

It should be emphasized that the velocity pulse, described in Section~\ref{subsec:velpulse}, is 
going to be activated in the middle of the second stage described in Section~\ref{subs:modelc}, when 
the clump has evolved up to 
time $t/t_{ff}=0.299$, when the peak density 
is $\log \left( \rho_{max}/\rho_0 \right)=0.23$. The pre-impact 
state of the clump is illustrated in Fig.~\ref{fCurf2}. 

All the models produce the formation of a ring of high density gas at the outermost region of 
the clump, whose physical properties, as width, mass and size depend obviously 
on the magnitude of the velocity $V_0$. As this dense gas collapses locally, then the 
ring will fragment, so that the relevant timescales are those of the fragments and not the global 
free-fall time of the clump. In spite of this, we will use $t_{ff}$ as the normalizing factor of the 
evolution time.     

\begin{figure}
\begin{center}
\includegraphics[width=4.0 in]{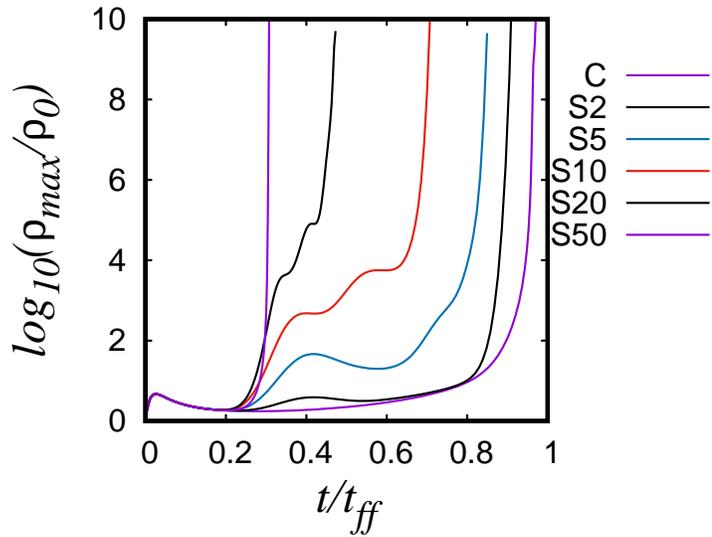}
\caption{\label{fDenMax} Time evolution of the peak density $\rho_{max}$ of Models C and S2--S50.}
\end{center}
\end{figure}

\begin{figure}
\begin{center}
\begin{tabular}{cc}
\includegraphics[width=3 in]{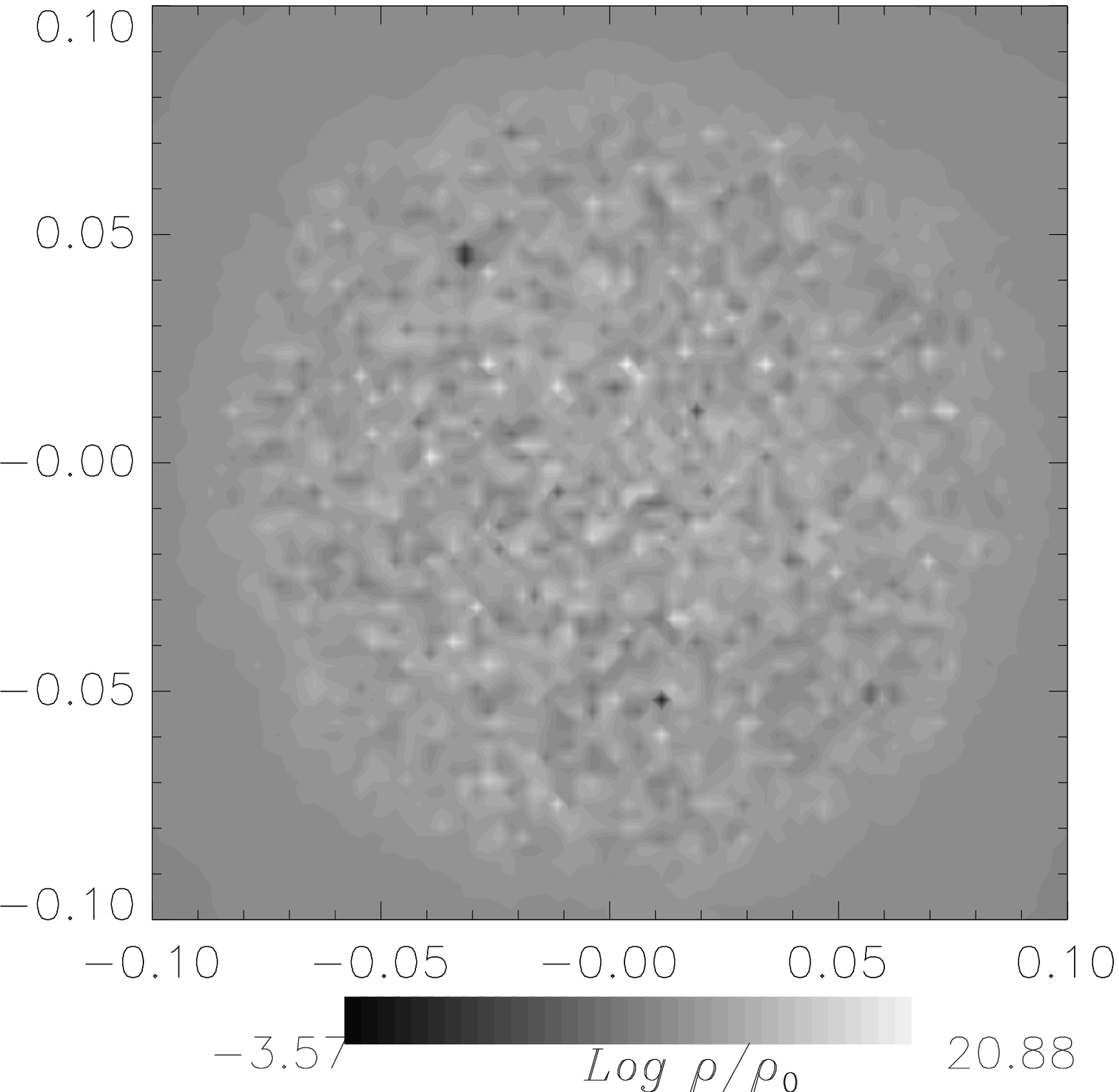} & \includegraphics[width=3.5 in,height=3 in]{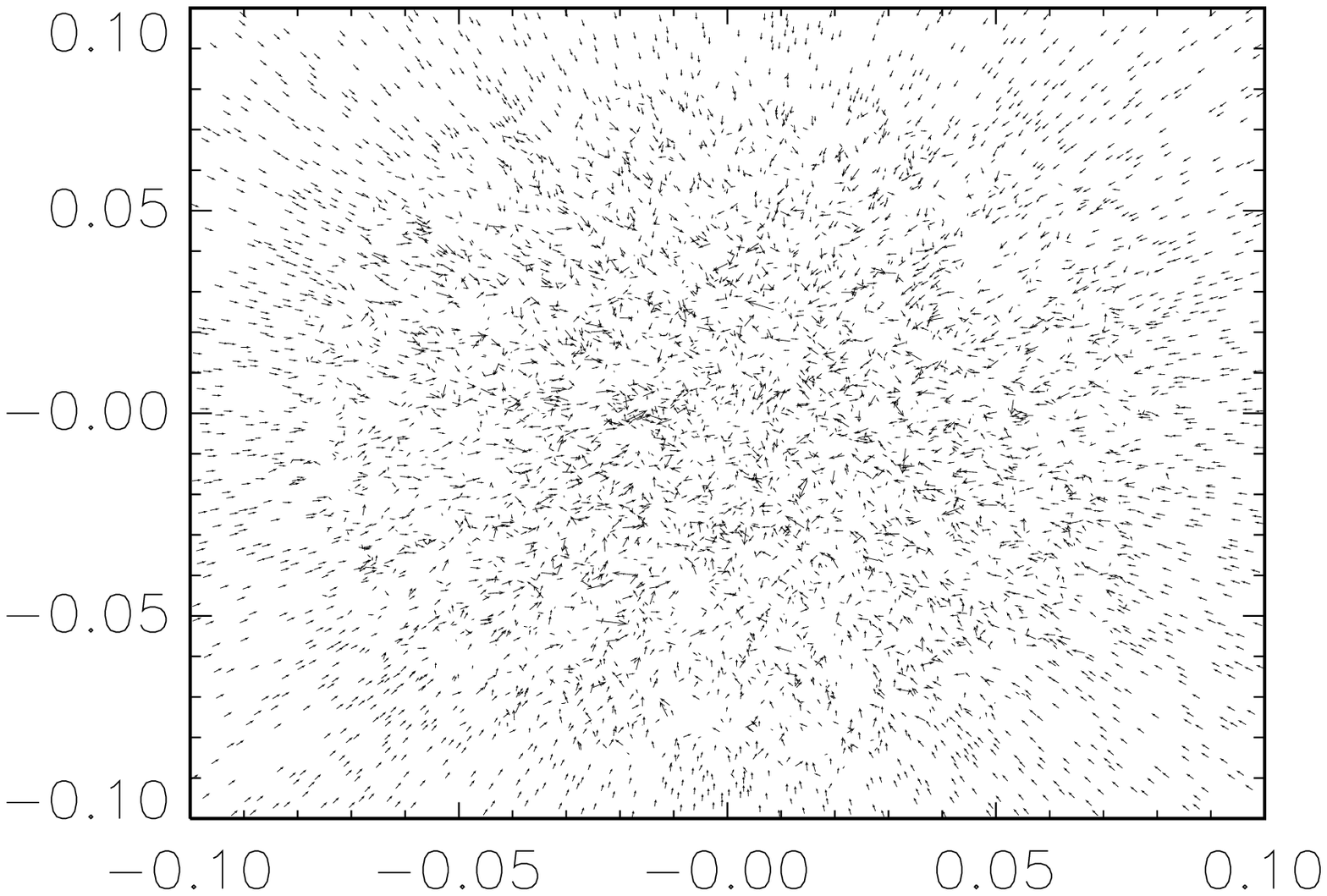} 
\end{tabular}
\caption{\label{fCurf2final} For the turbulent clump model at time $t/t_{ff}=0.97$ when the peak density 
is $\log_{10}\left(\rho_{max}/\rho_0\right)=16.5$, corresponding to the last snapshot obtained, the 
region $(-0.1 \, \times R_0, 0.1 \, \times R_0)$ of the X-Y midplane of the clump is shown 
by means of (left) a iso-density plot and (right) a velocity plot.}
\end{center}
\end{figure}

\begin{figure}
\begin{center}
\begin{tabular}{cc}
\includegraphics[width=3 in]{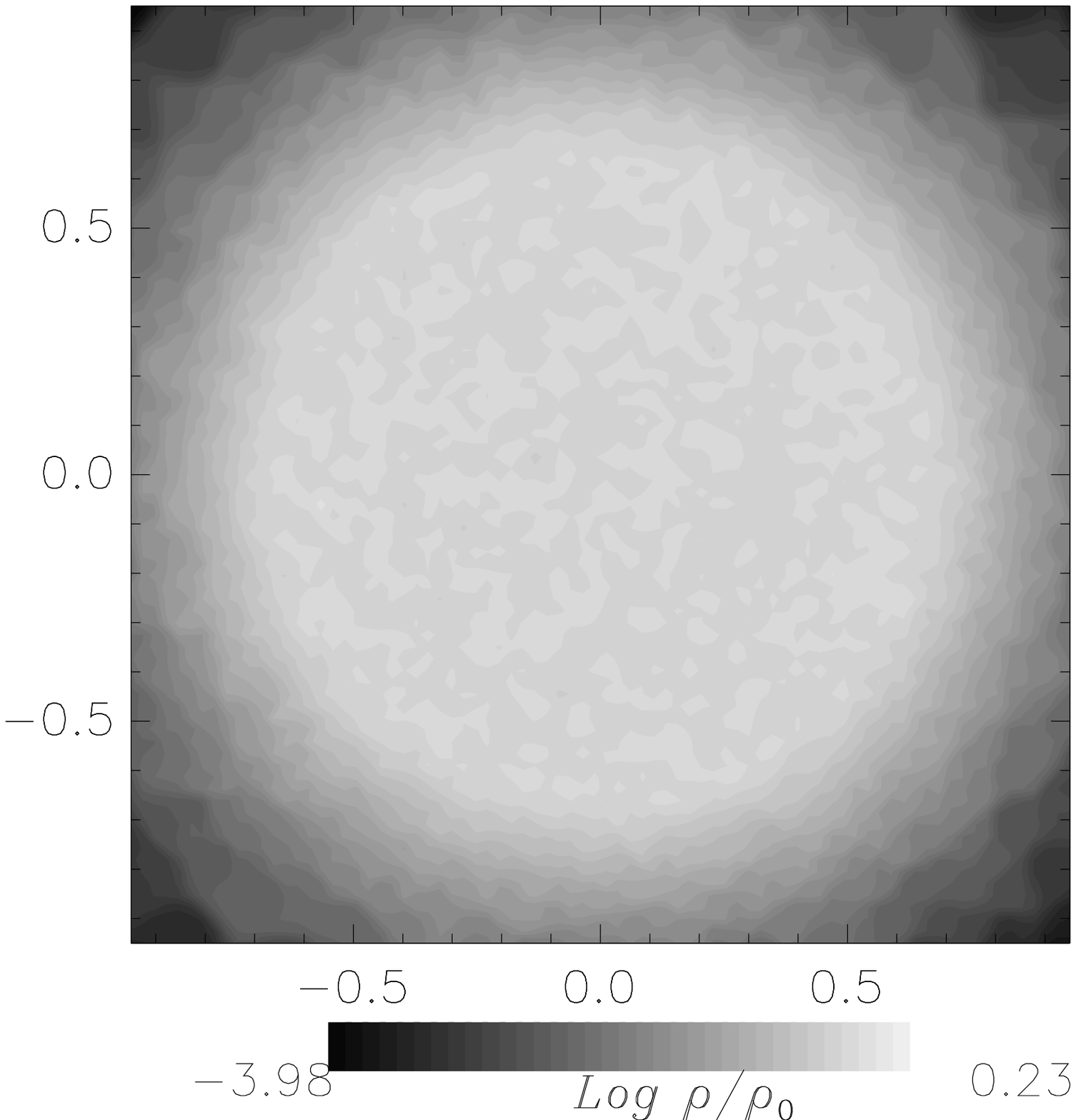} & \includegraphics[width=3.5 in,height=3 in]{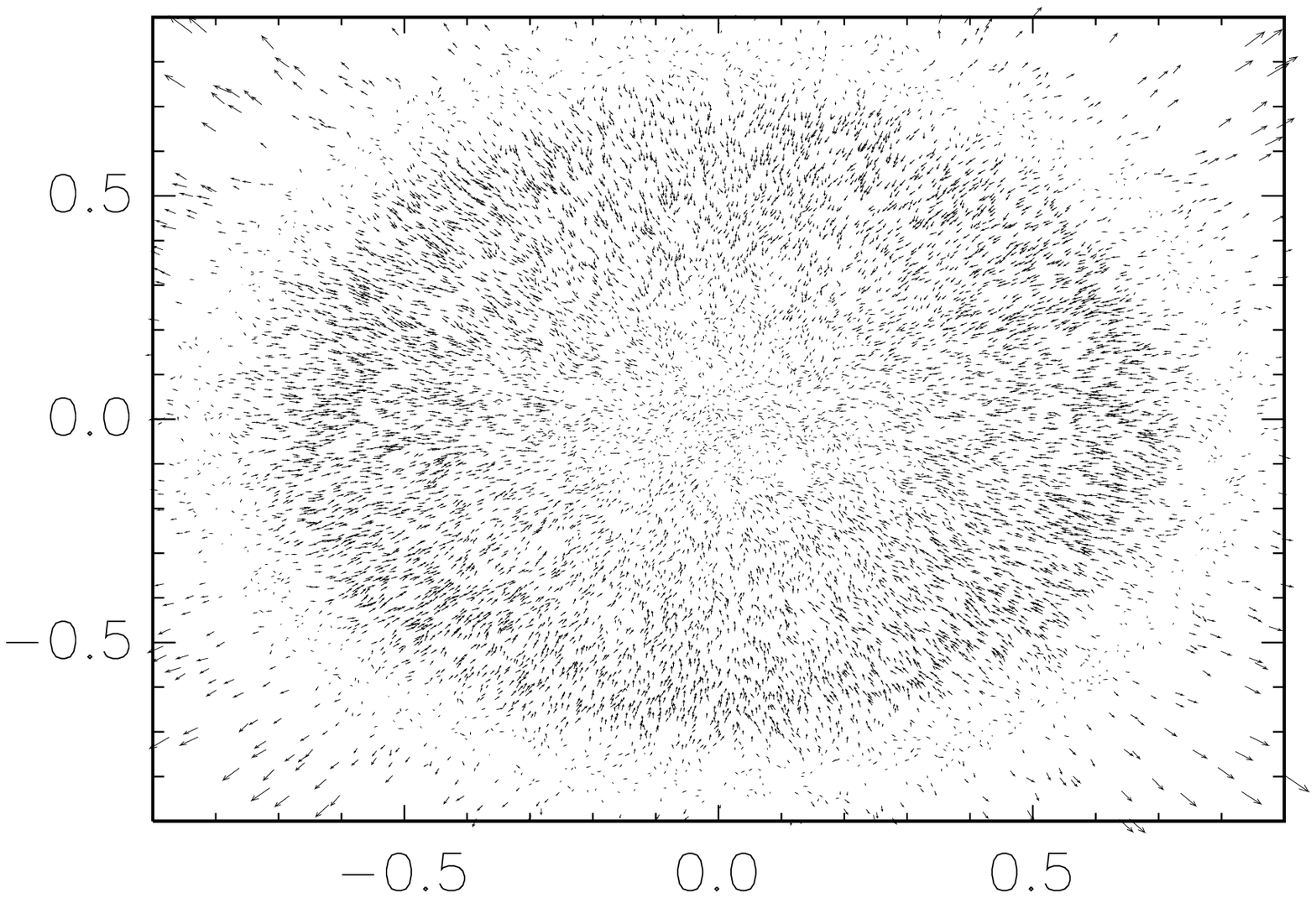} 
\end{tabular}
\caption{\label{fCurf2} For the turbulent clump model at time $t/t_{ff}=0.299$, when the peak density 
is $\log_{10}\left(\rho_{max}/\rho_0\right)=0.23$; this is the  moment at which the velocity pulse is activated 
in models S2-S50, the region $(-0.9 \, \times R_0, 0.9 \, \times R_0)$ of the X-Y midplane of the clump is shown 
by means of (left) a iso-density plot and (right) a velocity plot.}
\end{center}
\end{figure}

It was mentioned that Fig.~\ref{fDenMax} shows that all the models considered in this paper 
collapse at a time $t_{\rm col}$ less than the free-fall time 
of the isolated clump, $t_{ff}$. Clearly, $t_{\rm col}$ depends on the Mach number 
of the incident velocity, related to $V_0$ and shown in Table~\ref{tab:mod}. It can be appreciated 
that the collapse time $t_{\rm col}$ varies in general from 0.3 to 0.9 $\times \, t_{ff}$.

Another feature to be appreciated in Fig.~\ref{fDenMax} is 
that the non-linear evolution of the clump is reached very quickly, as follows. When the 
velocity pulse is activated in the gas clump, the Rankine--Hugoniot jump 
conditions, which analytically predict an increase of pre-shock density on the order 
of 4, is clearly surpassed, hereby the density oscillations in the curves of Fig.~\ref{fDenMax} are of 
order 10--100 times the average density of the clump.

Lastly, it should be mentioned that the models terminate their evolution 
because the huge increase in density makes the time step of the runs 
extremely small, to the point that the further evolution of the models 
can not be followed.       

In the subsequent sections \ref{subs:modelS2}--\ref{subs:modelS50}, the details of 
the resulting configuration of each shock model will be described separately. To complement this 
description, in Section \ref{sec:dist} an inter-model comparison will be presented. 
%%%%%%%%%%%%%%%%%%%%%%%%%%%%%%%%%%%%%%%%%%%%%%%%%%%%%%%%%%%%%%%%%%%%%%%%%%%%%  
\subsubsection{The model S2}
\label{subs:modelS2}

When the velocity $V_0$ takes its smallest value, $2 \, c_0$, for Model S2, the  
shell of dense gas is very thin initially. By the time $0.75 \, t_{ff}$, the shock reaches 
the interior region 
of the clump and a lower density of particles remain at the outer layers of the clump, so 
that the shell can hardly be distinguished.

\begin{figure}
\begin{center}
\begin{tabular}{cc}
\includegraphics[width=3 in]{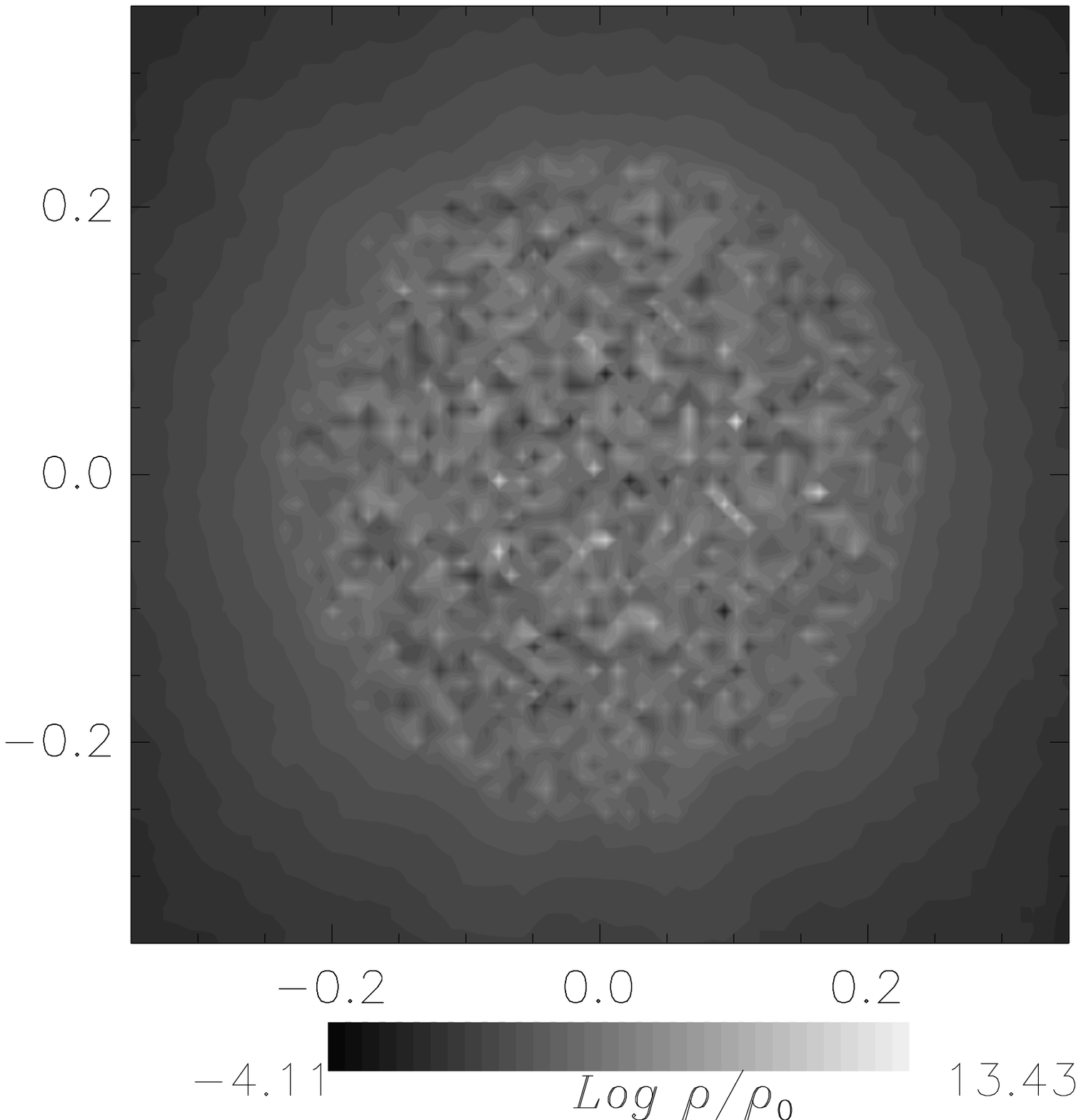} & \includegraphics[width=3 in]{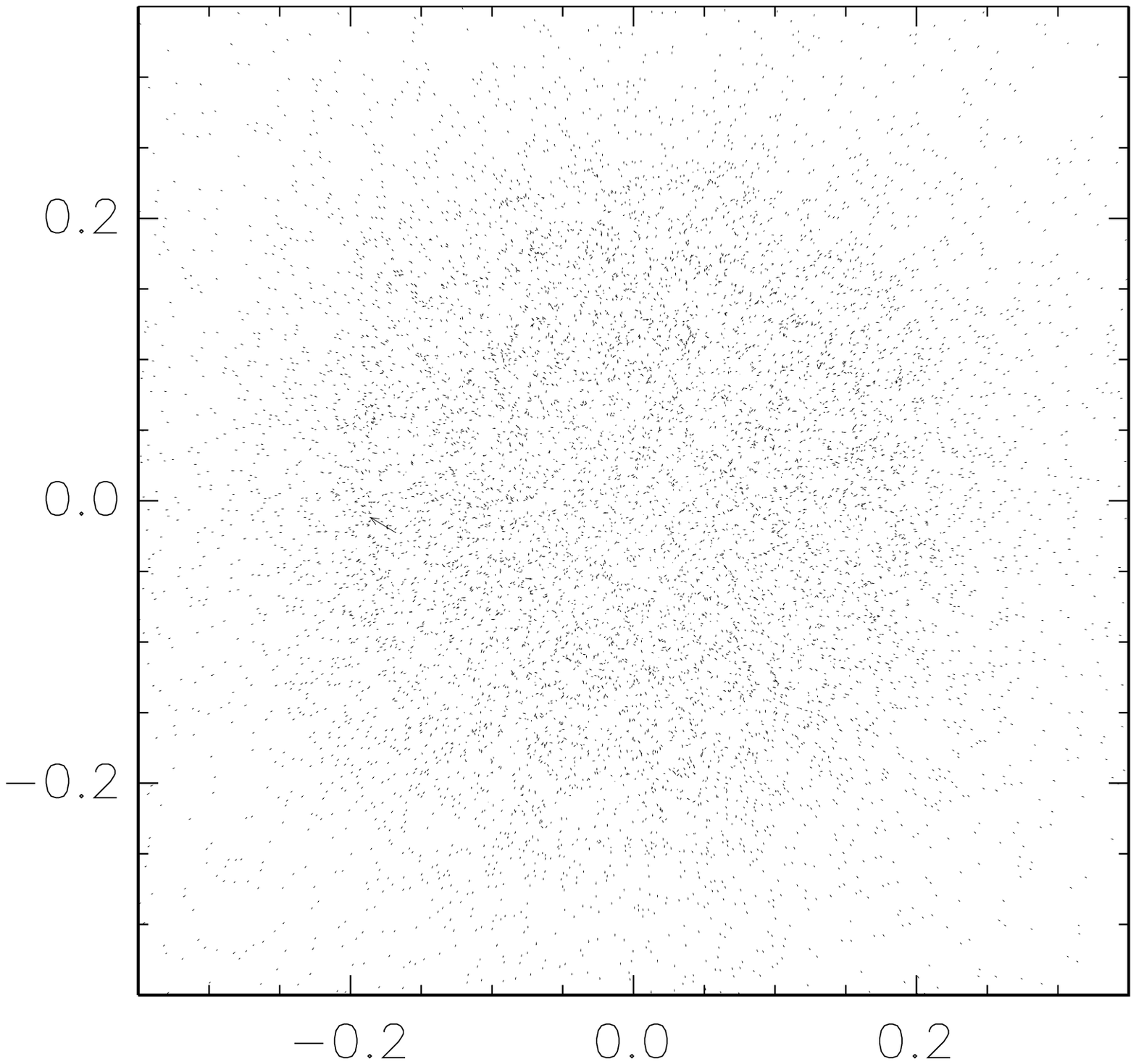} 
\end{tabular}
\caption{\label{fCurf2Sm2} When the velocity pulse is activated with initial velocity $2 \, c_0$, at 
time $t/t_{ff}=0.62$ when the peak density is $\log_{10}\left(\rho_{max}/\rho_0\right)=13.4$, the 
region $(-0.3 \, \times R_0, 0.3 \, \times R_0)$ of the x-y midplane of the clump is shown 
by means of (left) a iso-density plot and (right) a velocity plot.}
\end{center}
\end{figure}

Meanwhile, a higher density of particles is formed into the innermost region of the clump, so that the 
collapse goes faster there, although the cluster structure is not destroyed, so that it looks similar to that 
observed in Model C (when no velocity pulse was ever activated); see Fig.~\ref{fCurf2Sm2}.

The set-up of this model is similar to that implemented by \cite{gomez2007}, as they
used only one value of the velocity pulse, given by $2 \, c_0$, and two values of the 
parameter $r_0$. In the present paper, only one value of the parameter $r_0$ has been 
used, which is the same as one of the parameters used by \cite{gomez2007}, so that it was 
the one that led to a collapsing cloud.
 
\subsubsection{The model S5}
\label{subs:modelS5}

\begin{figure}
\begin{center}
\begin{tabular}{cc}
\includegraphics[width=3 in]{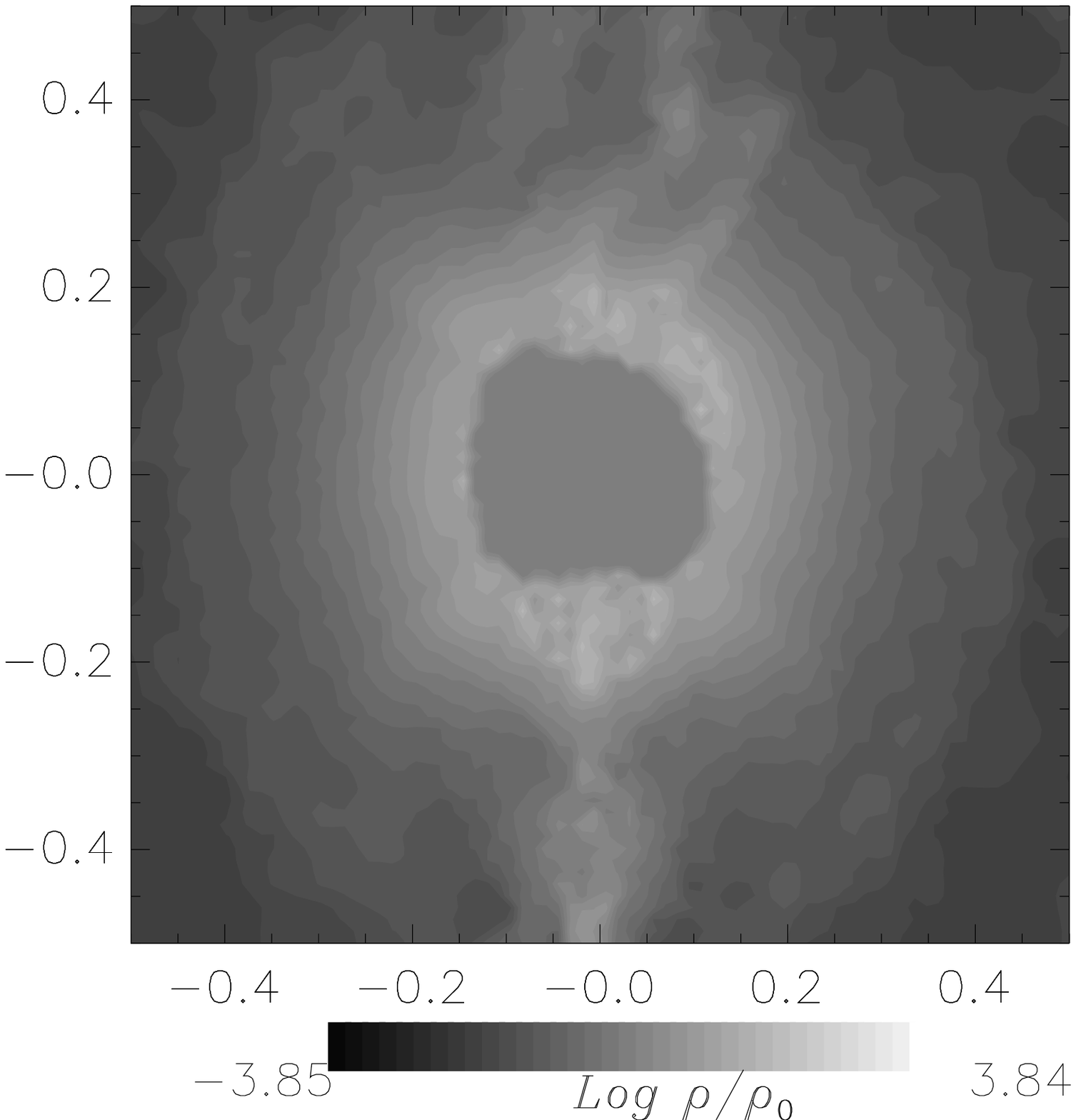} & \includegraphics[width=3 in]{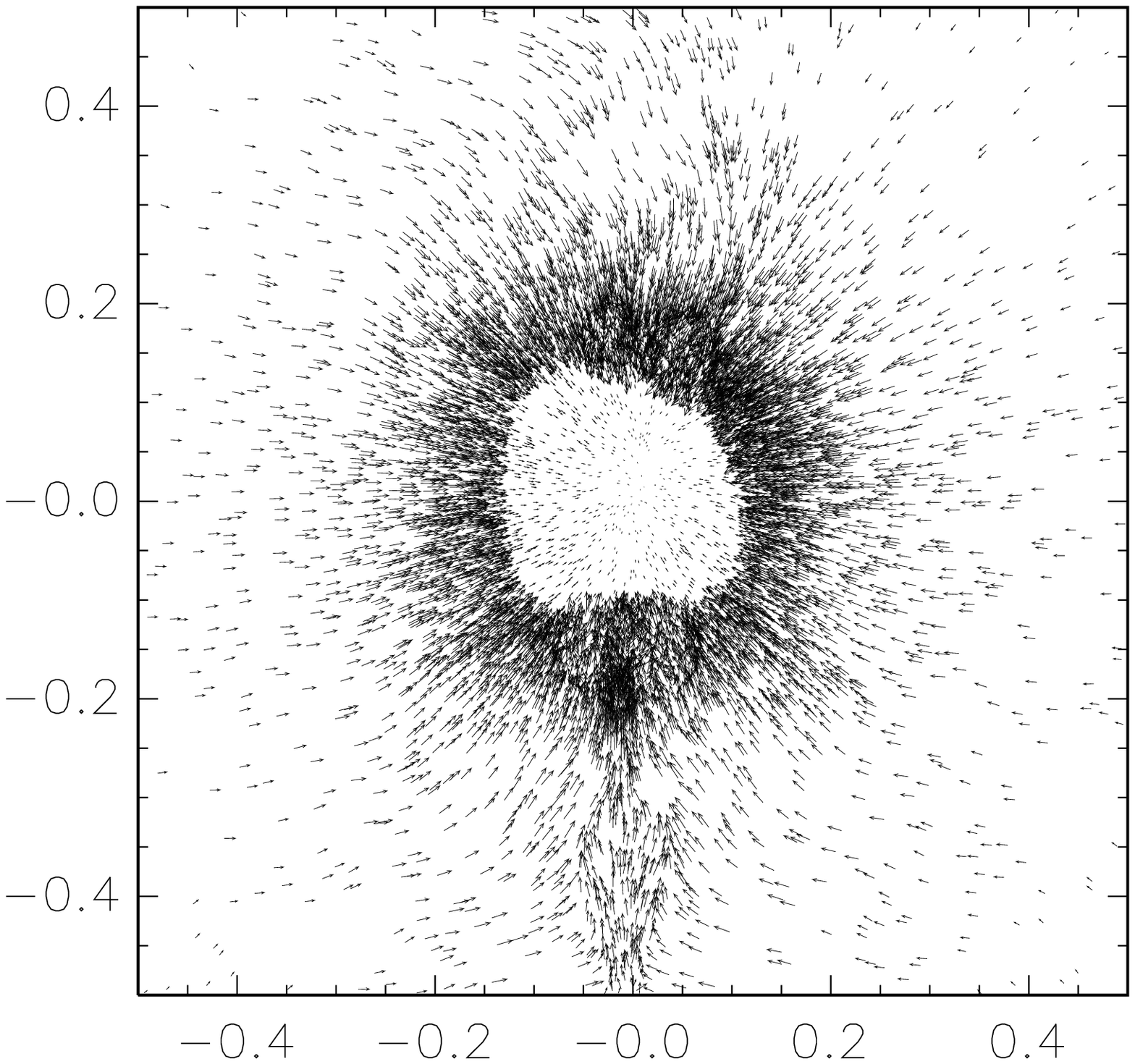} 
\end{tabular}
\caption{\label{fCurf2S5} When the velocity pulse is activated with initial velocity $5 \, c_0$, at 
time $t/t_{ff}=0.53$ when the peak density is $\log_{10}\left(\rho_{max}/\rho_0\right)=6.3$, the 
region $(-0.4 \, \times R_0, 0.4 \, \times R_0)$ of the x-y midplane of the clump is shown 
by means of (left) a iso-density plot and (right) a velocity plot.}
\end{center}
\end{figure}

In Model $S5$ with $V_0=5 \, c_0$, a dense shell of particles can clearly be seen.
It moves towards the centre of the clump and grows in mass while keeping its spherical 
symmetry. At the time $0.68 \, t_{ff}$, two outward asymmetric flows of gas occur 
simultaneously, one along the positive and the other along the negative direction of 
the $y$-axis, as shown in Fig.~\ref{fCurf2S5}. By the end of the 
simulation, this high density shell of particles is clearly visible in the innermost central 
region of the clump, so that the gas inside this enclosed region has a lower density of particles.           

\subsubsection{The model S10}
\label{subs:modelS10}

\begin{figure}
\begin{center}
\begin{tabular}{cc}
\includegraphics[width=3 in]{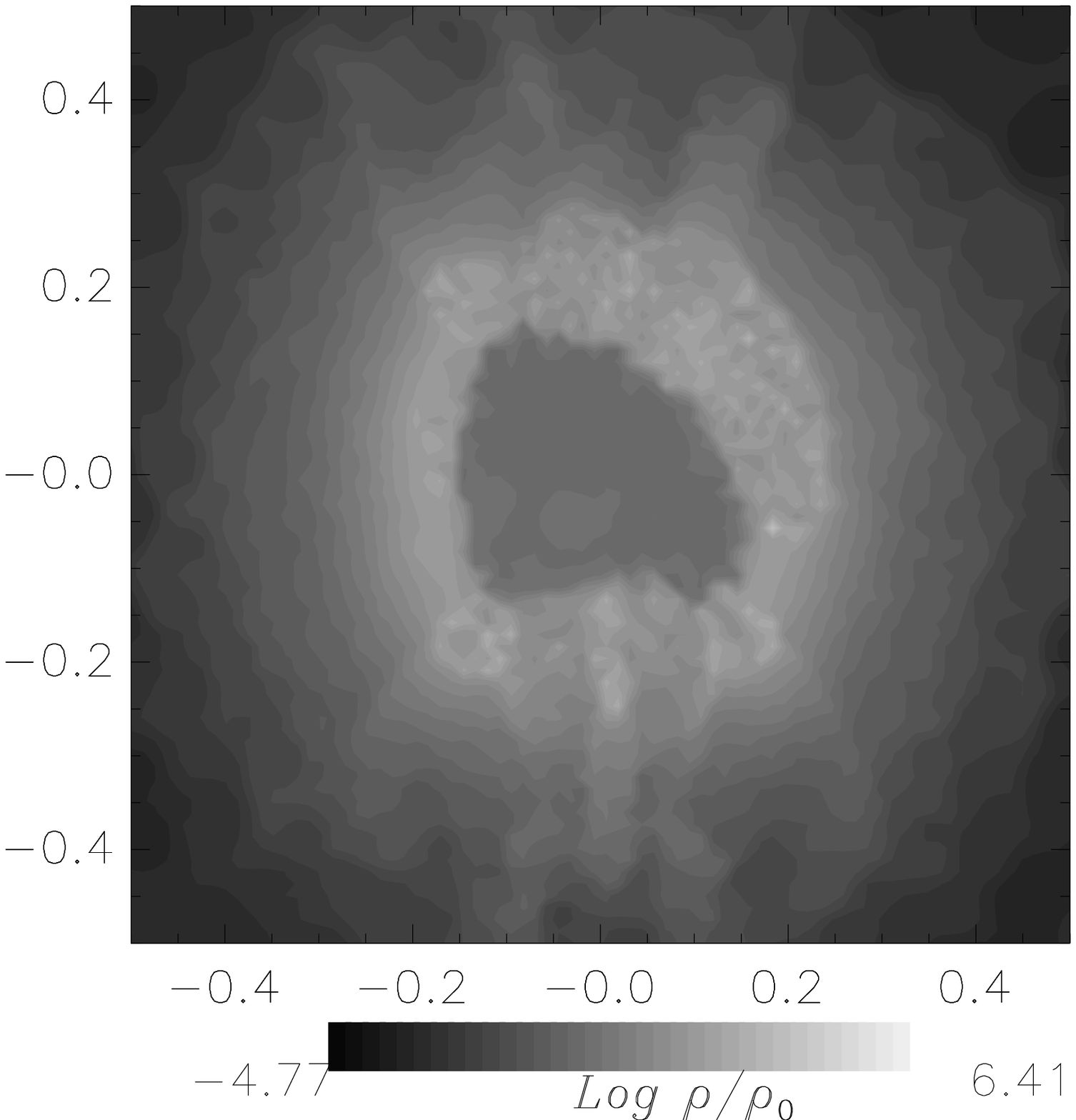} & \includegraphics[width=3 in]{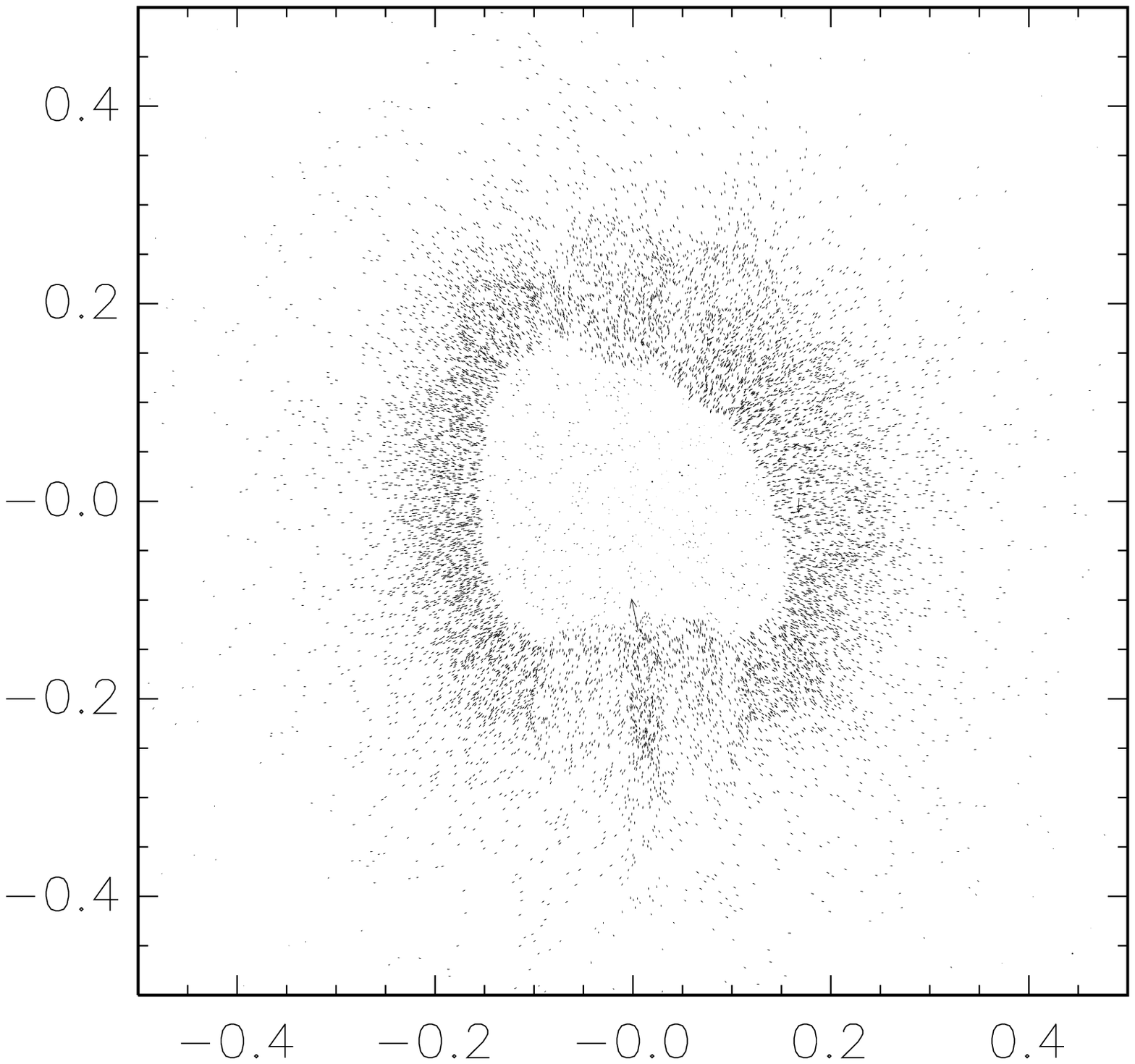} 
\end{tabular}
\caption{\label{fCurf2S10} When the velocity pulse started at $10 \, c_0$, at time $t/t_{ff}=0.44$ when the peak 
density is $\log_{10}\left(\rho_{max}/\rho_0\right)=6.4$, the region $(-0.4 \, \times R_0, 0.4 \, \times R_0)$ of the x-y midplane 
of the clump is shown by means of (left) a iso-density plot; (right) a velocity plot.}
\end{center}
\end{figure}

When the velocity pulse increases to $V_0=10 \, c_0$, many small protuberances 
appear along the dense shell of particles, so that the shell expels gas 
by means of these protuberances. As happened in previous model, in this case, 
at the time $0.47 \, t_{ff}$, much more gas accumulates simultaneously along the 
positive and negative directions of the $y$-axis, trying to leave outward. The dense shell of 
particles moves inward, so that it becomes very thick and encloses a very small 
region inside; see Fig.\ref{fCurf2S10}.  

\subsubsection{The model S20}
\label{subs:modelS20}

The evolution of Model S20, which has $V_0=20 \, c_0$, is very similar to that 
of Model S10. In this case, the gas protuberances are larger, so that they even show 
some curvature. In spite of this, the dense shell of particles does not show any sign 
of fragmentation. 

\begin{figure}
\begin{center}
\begin{tabular}{cc}
\includegraphics[width=3 in]{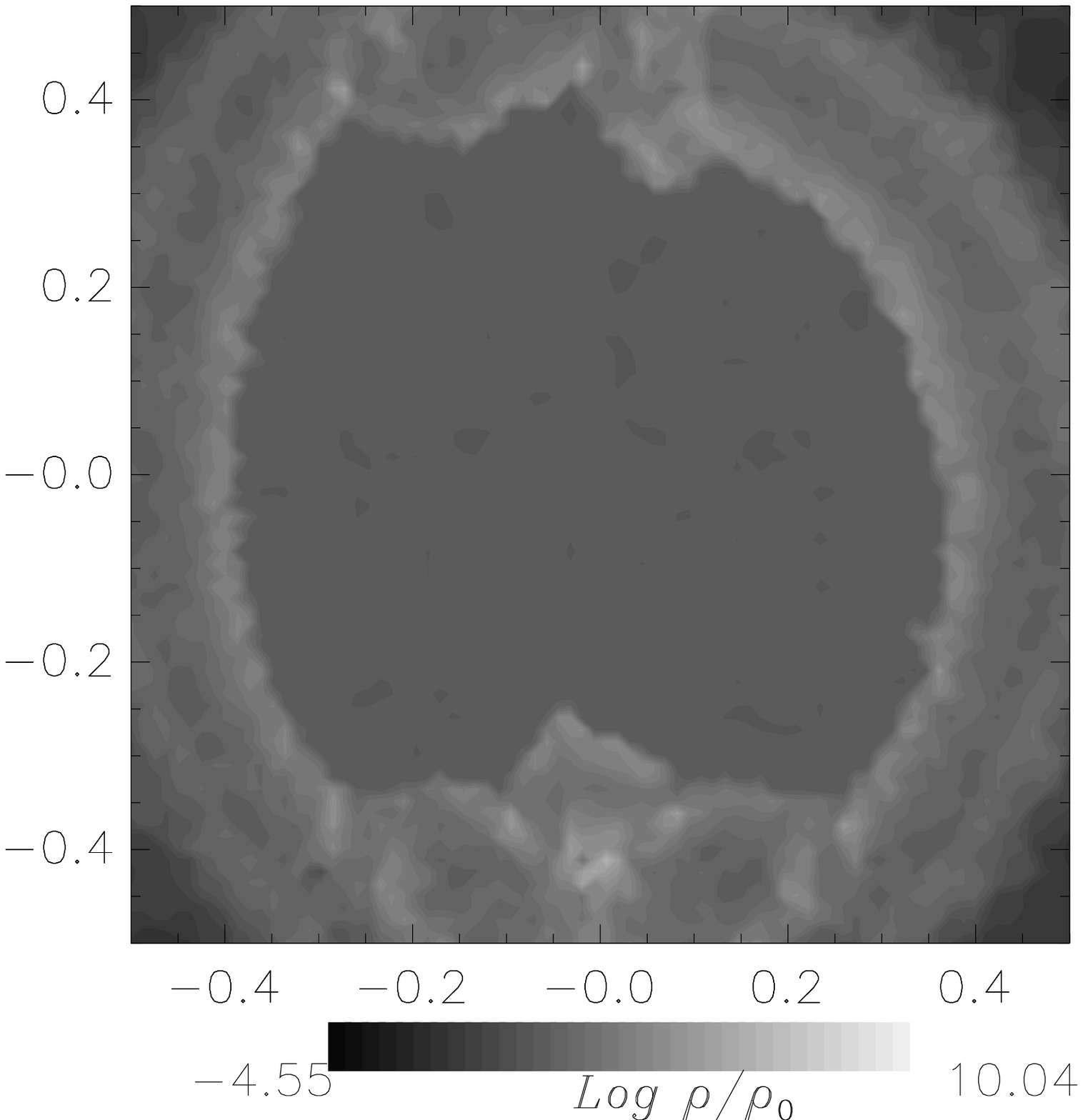} & \includegraphics[width=3 in]{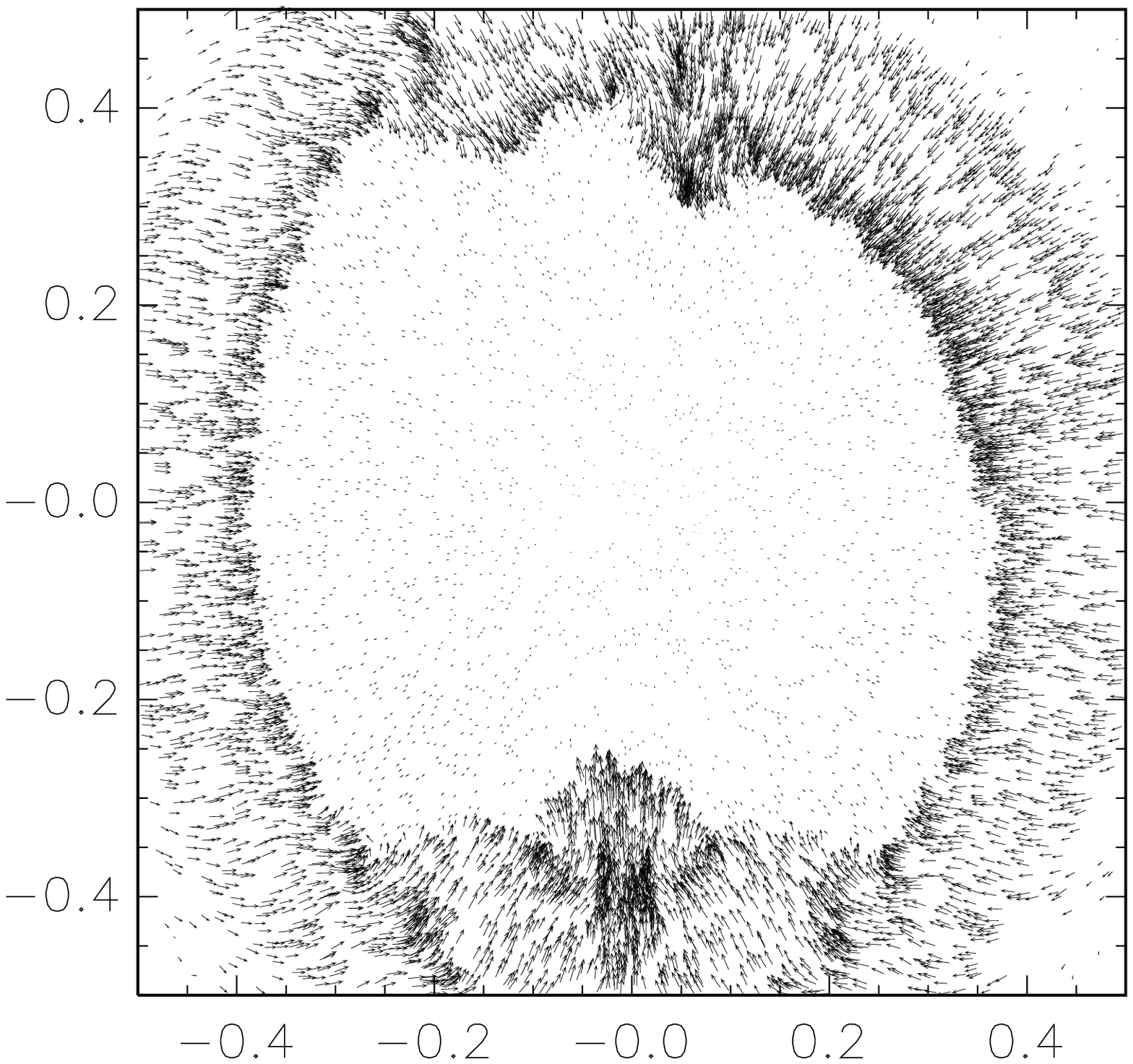} 
\end{tabular}
\caption{\label{fCurf2S20} When the velocity pulse started at $ 20 \, c_0$, at time $t/t_{ff}=0.17$ when the peak 
density is $\log_{10}\left(\rho_{max}/\rho_0\right)=10$, the region $(-0.4 \, \times R_0, 0.4 \, \times R_0)$ of the x-y midplane of 
the clump is shown by means of (left) a iso-density plot and (right) a velocity plot.}
\end{center}
\end{figure}

There is an outflow of dense gas through the negative 
direction of the $y$-axis, which is more noticeable than in the previous models. 
The simulation ends quickly, because the peak density is reached in the shell, without giving 
time for the shell to move inward appreciably, as was the case in Model S10; see Fig.~\ref{fCurf2S20}.

\subsubsection{The model S50}
\label{subs:modelS50}

\begin{figure}
\begin{center}
\begin{tabular}{cc}
\includegraphics[width=3 in]{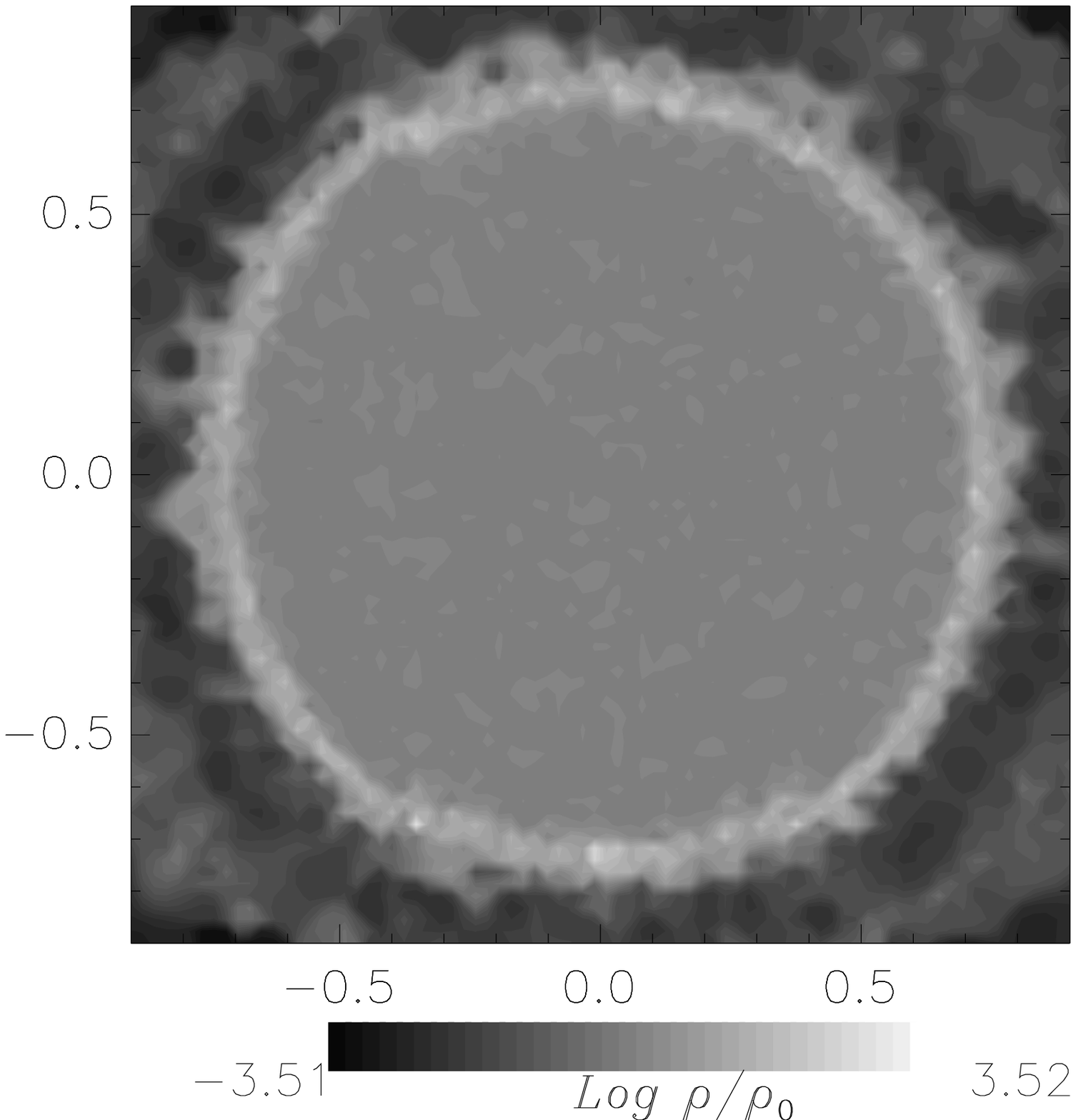} & \includegraphics[width=3 in]{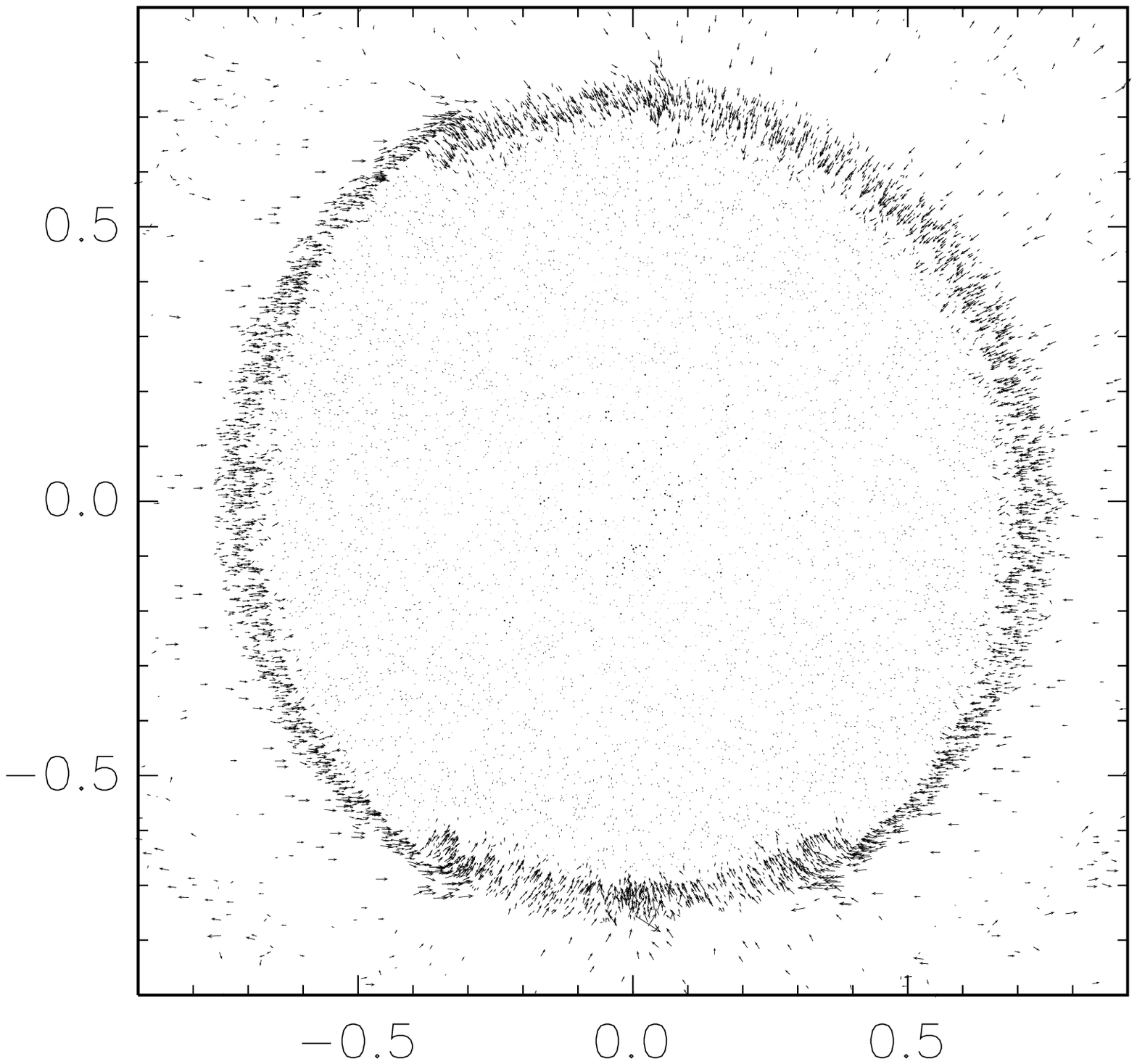} 
\end{tabular}
\caption{\label{fCurf2S50} When the velocity pulse started at $50 \, c_0$, at time $t/t_{ff}=0.01$, when the peak 
density is $\log_{10}\left(\rho_{max}/\rho_0\right)=3.5$, the region $(-0.9 \, \times R_0, 0.9 \, \times R_0)$ of the x-y midplane 
of the clump is shown by means of (left) a iso-density plot and (right) a velocity plot.}
\end{center}
\end{figure}

In Model $S50$, for which the velocity $V_0$ takes the largest value considered in the present 
paper, the dense shell of particles is very thin. The increase of density 
after the shock is high enough to favour the fragmentation of the shell in many places 
throughout the shell simultaneously. However, as more particles of the clump envelope are 
still falling into the shell, it keeps its spherical symmetry. The simulation ends 
because some of the fragments of the shell reach their peak density so quickly, that there was 
no time for the shock to move anymore; see Fig.~\ref{fCurf2S50}. 
   
%%%%%%%%%%%%%%%%%%%%%%%%%%%%%%%%%%%%%%%%%%%%%%%%%
\section{The radial profile of the density and velocity distributions}
\label{sec:den2}

In order to characterize further the distribution of particles, their dynamics and the dense shell of 
particles in all the shock models considered above, in this section a radial partition 
of the spherical model is performed as follows. 

\begin{figure}
\begin{center}
\begin{tabular}{cc}
\hspace{-1 cm} \includegraphics[width=4 in]{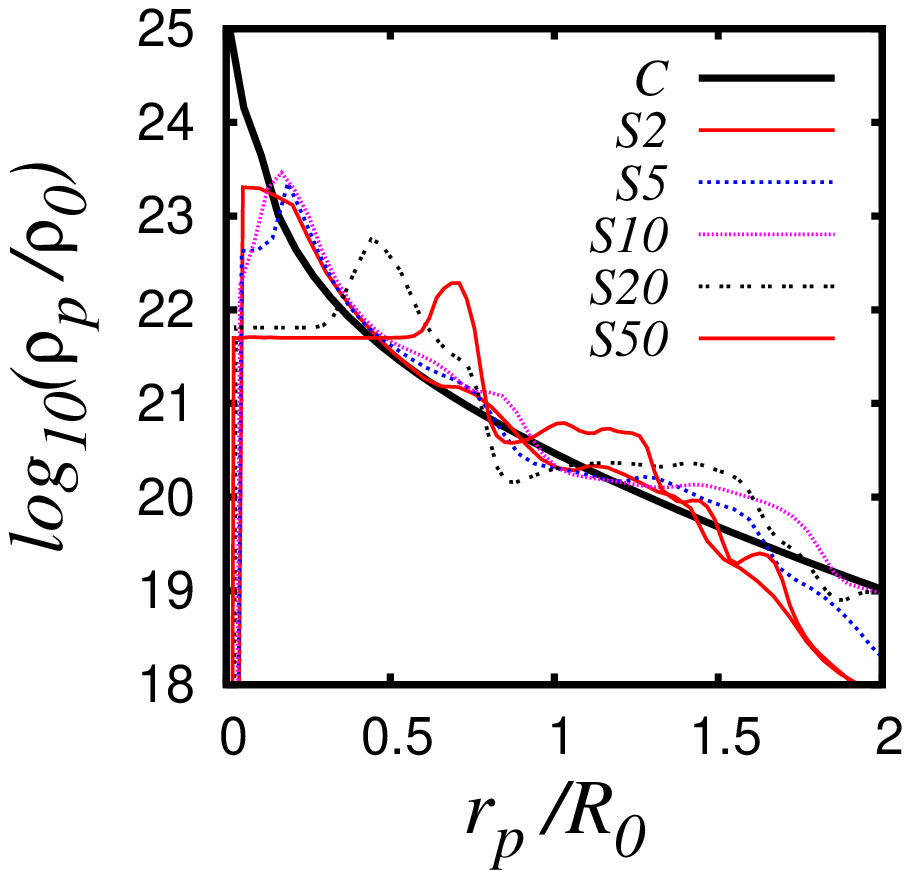} & \hspace{-2 cm} \includegraphics[width=4 in]{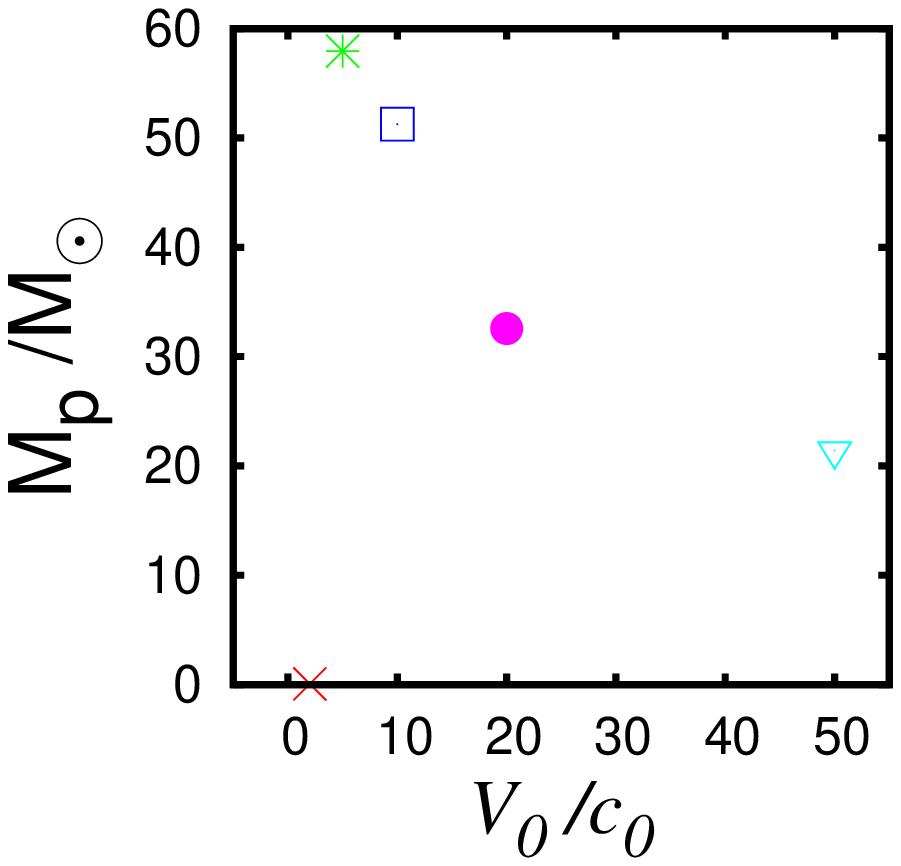} 
\end{tabular}
\caption{\label{fMasar} (left) The log of the ratio between the radial density to the average initial 
density, against the radius of the sphere model normalized with the initial clump 
radius $R_0$; (right) the mass contained in the dense shell formed at the last snapshot available 
for each model, shown here in the x-axis.}
\end{center}
\end{figure}

The maximum radius $r_{\rm max}$ to which the spherical clump has expanded by 
the end of its evolution time, has been determined for each model. Then, a radial partition of the spherical 
clump in $n_{\rm bin}$ bins is made from 0 to $r_{\rm max}$, so that the increments in radius are 
given by $\delta r=r_{\rm max}/n_{\rm bin}$; thus, all the particles contained in a 
radial bin within the radii $r_i$ and $r_i +\delta r$ are taken into account 
to calculate their average particle density, which is shown at radius $r_p =r_i +\delta r /2$ 
in the left panel of Fig.~\ref{fMasar}. It should be noted that the density $\rho_p$ and radius $r_p$ are 
normalized with the initial density and radius of the isolated clump, $\rho_0$ and $R_0$, 
respectively.   

Several interesting observations can be made from the left panel of Fig.~\ref{fMasar}. First, it clearly 
shows that the greater the Mach number of the incident shock (the $V_0$ of the model), the further 
away from the clump centre is formed the shell of 
dense gas. Second, it also shows that the larger the initial 
velocity of the pulse $V_0$, the lower the central density. Third, all the curves of the shock 
models follow the 
radial density profile of the isolated clump. Fourth, as expected on the basis of the 
Rankine--Hugoniot jump conditions, the oscillations in the density curve are more 
pronounced in the case of models with a large pulse velocity. 

Taking advantage of this radial partition, the average velocity of the particles 
is calculated per radial bin, so that the result for the first snapshot was already shown in 
Fig.~\ref{velCurf2S}, while the result for the last snapshot available in each model is now presented in 
Fig.~\ref{velCurf2SFinal}. Consider now Fig.~\ref{velCurf2SFinal}. In the case of the clump 
model, the outer particles of the central region continue to fall towards the centre of the 
clump. This is also observed to occur in the low velocity $V_0$ shock models, namely, Models S2, S5 and S10. However, for 
the higher velocity $V_0$ shock models, namely Models S20 and S50, the central region shows no particles moving at all, until the 
radius at which the first dense shell is reached.

A velocity field similar to the one described above for Models S20 and S50 was observed by \cite{henne} in the cases of 
strongly supersonic, finite compression, where the external pressure on the core increased until it was halted once it 
was doubled, see his model with $\phi=0.1$.     
  
%%%%%%%%%%%%%%%%%%%%%%%%%%%%%%%%%%%%%%%%%%%%%%%%%
\section{The mass of the shell}
\label{sec:themass}

To calculate the mass contained in the dense shell, denoted 
by $M_p$, we apply again the radial partition of the spherical model described in 
Section \ref{sec:den2} to the last snapshot available\footnote{Unfortunately, the masses calculated in this way, are 
going to be obtained for quite different evolution times; in spite of this, their comparison is interesting as it 
allows us to characterize quantitatively the last configuration available.}, so that the radial bin at which the 
highest density value takes place must be now 
firstly obtained. In this case, a variation of density is also needed to define the width of 
the shell; let it be defined as $\delta \rho_p$ with a value of 
$\log_{10}\left( \delta \rho_p \right) = $ 0.5 for all the 
models. Thus, all those particles with a density 
within the range $\left( \rho_p -   \delta \rho_p, \rho_p +  \delta \rho_p \right)$ and located within the 
radii $\left( r_p - 2  \delta l, r_p + 2  \delta l \right) $ are selected to obtain the mass of 
the shell. The parameter $\delta l$ has been chosen visually from Figs~\ref{fCurf2S5}--\ref{fCurf2S50}, so that 
there is one value for each shock model as follows: 0.03 for model S2; 0.06 for models 
S5 and S10; 0.033 for model S20 and 0.016 for model S50. The result of this calculation can be seen in the right panel 
of Fig.~\ref{fMasar}, against the the initial velocity of the shock model. 

It is interesting to note 
that the larger the initial velocity $V_0$ of the pulse, the lower the mass $M_p$ contained in the 
shell.  However, it is remarkable that the mass contained in the shell is in general 
very large, so that the ratio of this mass to that of the clump ranges within 0.32--0.1.

%%%%%%%%%%%%%%%%%%%%%%%%%%%%%%%%%%%%%%%%%%%%%%%%%%%%%%%%%%%%%%%%%%%%%%%%%%%%%%%%%%%%%%%%%%%%%%%%%%%%%%%%%%
\section{Time evolution of the distance of the shell to the clump centre}
\label{sec:dist}

It was mentioned in Section \ref{sec:results} that the models were advanced in time as much 
as possible, up to the time when the time-step becomes prohibitively small for the simulations to evolve 
any further, so that in Sections \ref{subs:modelS2}-\ref{subs:modelS50} the last snapshots were chosen 
to illustrate the simulations outcome. 

In this Section, an attempt is made to make an inter-model comparison, by calculating the distance of 
the dense shell of gas to the clump centre for the shock models, as follows. In Sections \ref{sec:den2} 
and \ref{sec:themass}, the peak density in a radial partition was determined in order to define the width of the dense 
shell of gas. In this section we consider again this peak density to locate the shell of dense gas with 
respect to the clump centre, for all the snapshots available of each shock model. The result of this 
procedure is shown in Fig.\ref{fDisCte}. 

When the shock front is activated, there is an immediate increase of density, so that 
the shell of dense gas is located initially very close to $r_0$, the initial radius in the mathematical function 
of the velocity pulse, see Eq.\ref{velpulse}. Then, the shell moves radially inward as can be seen 
in Fig.\ref{fDisCte}. Unfortunately, the evolution time reached for model S50 is quite small; for model S20, the 
shell of dense gas was followed up to a radius around $0.43 \times R_0$ from the clump centre. For models 
S10 and S5, the closest approaching distance to the clump centre was around $0.15 \times R_0$. The most 
interesting behaviour was observed for model S2, since there is a bounce of the shell of dense gas. 

The horizontal dashed line shown in Fig.\ref{fDisCte} indicates the radius with which  
the snapshots for this inter-model comparison are now determined, since all the snapshots 
must have their shell of dense particles at an equal distance to the clump centre, given by this radius, as 
can be seen in the iso-density plots shown in Fig.\ref{fMos4}.   
%%%%%%%%%%%%%%%%%%%%%%%%%%%%%%%%%%%%%%%%%%%%%%%%%%%%%%%%%%%%%%%%%%%%%%%%%%%%%%%%%%%%%%%%%%%%%%%%%%%%%%%%%%%%
\section{Discussion}
\label{sec:dis}

The pulse velocities considered in this paper, the $V_0$ shown in Table~\ref{tab:mod}, both in terms 
of the speed of sound and in the MKS system of units, are 
comparable with those used in the following papers: (i) \cite{vanhala}, who considered the range 
10 to 50 km/s; (ii) \cite{boss95}, who used the range 0 to 25 km/s; and (iii) 
\cite{gomez2007}, who considered only one velocity given by twice their 
speed of sound, which corresponds to 0.4 km/s. 

It should be mentioned that the papers of \cite{vanhala} 
and \cite{boss95} are not directly comparable with the present paper, as they considered planar shock waves, not 
spherical velocity pulses.  \cite{vanhala} defined a critical shock velocity of 20 km/s,
which is required to obtain a triggered collapse of their spherical, rotating 
core. For shock velocities above it, around 45 km/s, the shocks are disruptive 
whereas for shock velocities below it, around (or less than) 10 km/s, the core 
rebounds and is torn apart. \cite{boss95} found that a shock wave striking 
a rotating core at a velocity of 25 km/s resulted 
in the `outside-in' collapse of the core. Both \cite{vanhala} and 
\cite{boss95} consider a rotating centrally condensed core, but it seemed 
that rotation did not make any significant difference in the outcomes of the simulations.  

In the case of the present paper, it was mentioned in 
Section~\ref{subs:modelc}, that the pulse was activated when 
the first stage of evolution of the clump has passed, so that the peak density 
does not increase significantly due to the random collisions between particles; thus, it is 
expected that the turbulence has reached at this time a fully developed state. This 
objective was also mentioned by \cite{vazquez2008}, who made a pre-run in their simulations, so 
that gravity was turned on only after 3.2 turbulent crossing times of evolution of 
the turbulent cloud.    
   
\begin{figure}
\begin{center}
\begin{tabular}{cc}
\hspace{-1 cm} \includegraphics[width=4.0 in]{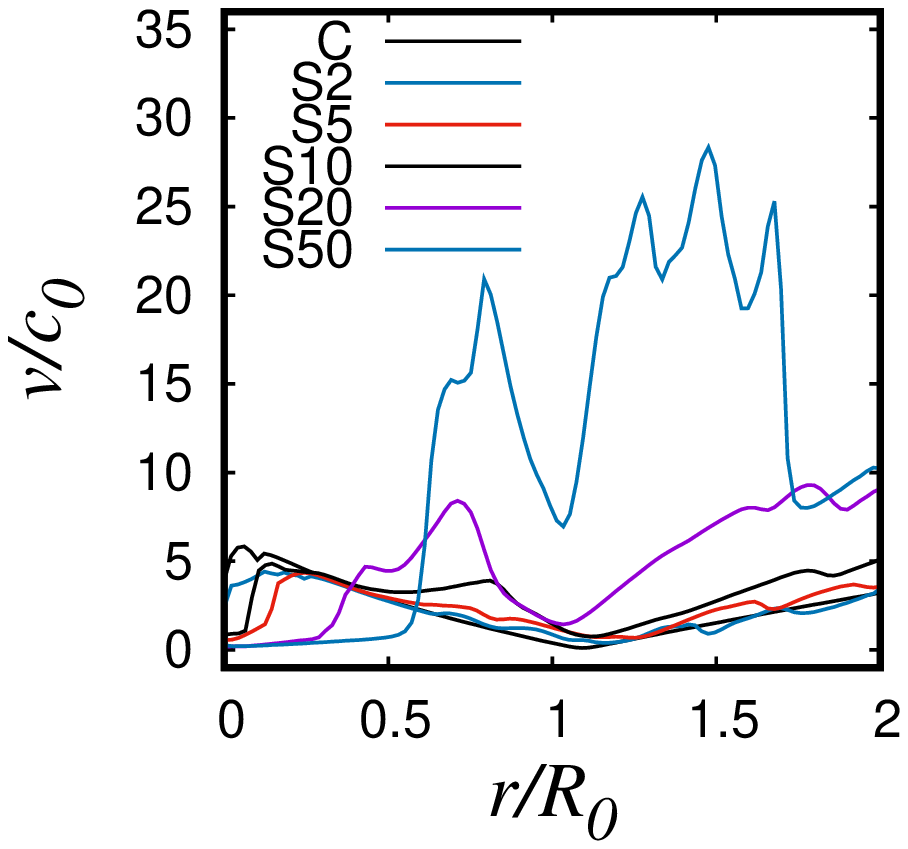} & 
\hspace{-3 cm} \includegraphics[width=4.0 in]{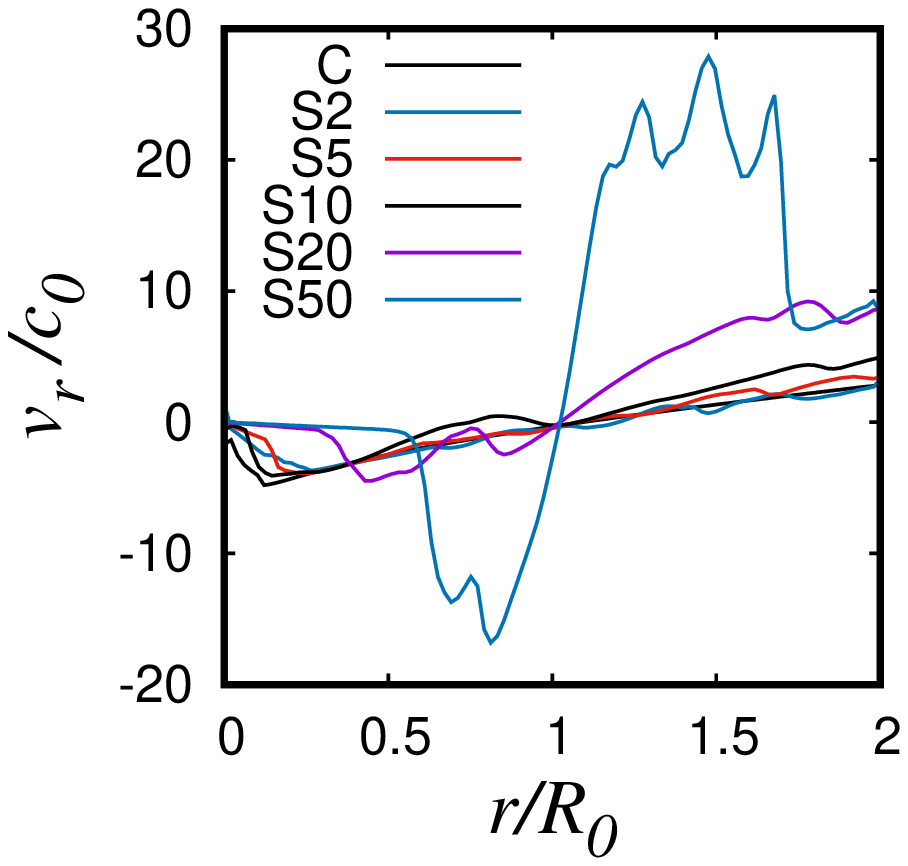}
\end{tabular}
\caption{\label{velCurf2SFinal} For the last snapshot available for each model, the measured velocity profile 
against the clump radius for Models S2--S50 is shown by plots of the (left) velocity magnitude and (right) radial 
projection of the velocity. Velocities are normalized with the initial speed of sound of the clump.}
\end{center}
\end{figure}

The gas particles with a radially inward initial velocity given directly by the velocity 
pulse, that is, those that are originally located around a radius $0.8\, R_0$, see 
Fig.~\ref{velCurf2S}, reach those particles located at smaller radii, giving place to the 
formation of the shell of dense particles observed in the present paper; this is a manifestation of 
the so-called `collect and collapse model', simulated by \cite{dale}. However, there is a significant difference 
between these two scenarios, with particles going inward or outward, as the boundary 
conditions imposed on the moving shock are different, that is, for the inward case, there 
must be a rebound of those particles that reach the centre of the clump (reflective boundary 
conditions); one of these effects can be seen in Models S5 and S10 of 
the present paper, in which the dense shell of particles is very deformed as it approaches the 
centre of the clump 
and even stops moving inward. In the outward case, the outflow continues without any 
boundary. It should be emphasized that on the contrary to the observation of \cite{vazquez2008}, the plot 
Fig.~\ref{velCurf2SFinal} does not suggest any evidence for the occurrence of a shock bouncing at 
the center of the shock models. However, in this paper a bounce was observed only for 
Model S2, see Section \ref{sec:dist}.  

It was mentioned in Section~\ref{sec:intro} that the infall velocities for the cores L1689B 
and L694-2 were observed to be faster than expected, see \cite{lee}.  
A theoretical model proposed by \cite{seo}, in which the collapse of the cores was modeled by means of 
a uniform density or as Bonner--Ebert spheres, demonstrated that these cores may have infall velocities 
up to -1.0 or -1.5 times the velocity normalized with the speed of sound, which indicates an enhanced collapse. 
Meanwhile, for the cores L1544 and L63, in which the collapse appeared to be normal, the magnitude of 
the infall velocity was around -0.5 times the speed of sound\footnote{The magnitudes of the infall velocities 
reported here have been taken from fig.~6 of \cite{seo}.}. These authors suggested that the 
former cores may be strongly influenced by an external factor. 

From the results obtained in the simulations considered in the present paper, such an external factor 
can not be a velocity pulse with the parameters chosen here. According 
to Fig.~\ref{fDenMax}, it can be observed that the overall 
collapse of the clump model was enhanced by the velocity pulse. However, the infall velocities of the low 
velocity shock models were observed to be a little bit smaller than those observed for the isolated clump model, 
see Fig.~\ref{velCurf2SFinal}. 

In addition, the observations of the core L1544 by 
\cite{tafalla} and \cite{williams} suggested that ''the inner parts of the core are relatively stationary, and
an approximately uniform velocity field has been established in the
outer layers.'' For the higher velocity $V_0$ shock models of this paper, namely Models S20 and S50, it can 
be very clearly seen that the central region of the clump shows no particles moving at all, until the 
radius at which the first dense shell is reached. The particle velocities in the inner region are 
essentially zero because of the short simulation time. It was also observed in Section \ref{sec:den2} that the clump 
separates into two halves, so that there is one half with particles falling inward while 
there is another half in the exterior region with particles flowing outward.  

Another feature to be remarked here is relative to considering the velocity pulse as a core formation 
mechanism, as was originally proposed by \cite{gomez2007}, see Section~\ref{sec:intro}. In fact, using 
one-dimensional radial simulations and only one initial velocity of the pulse given by 2 
times the speed of sound, \cite{gomez2007} observed that once the shock reaches and bounces off 
the centre of their cloud, a central core is formed which is surrounded by a gas 
envelope. 

In the present paper, three-dimensional simulations of a set-up similar to that 
proposed by \cite{gomez2007} have been carried out. It should be mentioned that a direct comparison of 
the results of this paper with  
with the work of \cite{gomez2007} can not be made, since their initial models were not prepared 
to collapse spontaneously. On the contrary, the dense central region in the present paper is formed even in 
the model without a velocity pulse, since our clump was prepared to collapse spontaneously.

In addition, in this paper several initial velocities of 
the pulse have been taken into account. The radial distribution of the density of the 
shock models shown in the left panel of Fig.~\ref{fMasar} can be compared with the first line of 
the plots shown in \cite{gomez2007} in their fig.~3. Moreover, in Fig.~\ref{velCurf2SFinal}, it can be seen that the 
low velocity models have radial velocity curves with a behaviour very similar to that 
described by \cite{gomez2007} in the third line of panels of their Fig.~3. 

Therefore, on the basis of this limited comparison, the 
finding of \cite{gomez2007} is here confirmed, that a dense central region is 
formed as a result of the shock--gas interaction. However, one prediction of the present simulations 
is that there is a critical initial velocity of the pulse such that for shock velocities 
below the critical velocity, a central dense region will be formed, but otherwise, the central region 
is almost empty of moving gas. As we mentioned in Section \ref{sec:den2}, a core configuration with a velocity field 
similar to the one described here, was obtained by \cite{henne} in the case of a core subjected to a strong 
external pressure.     

A second prediction of this paper is that the central region strongly resembles the 
configuration obtained from the collapse of the isolated gas structure, without the velocity 
pulse, which in this case was obtained only for the smallest shock velocity, that 
of twice the speed of sound. 

\begin{figure}
\begin{center}
\includegraphics[width=4.0 in,height=3.0 in]{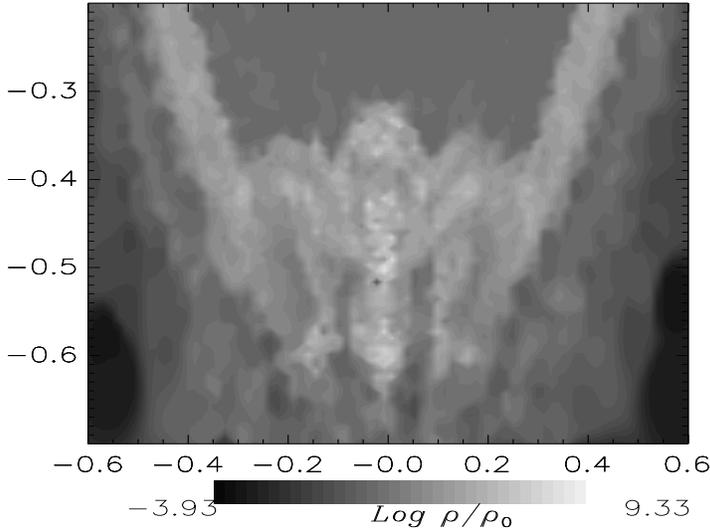}
\caption{\label{fCurp2pS20Zi} A zoom-in of the outflow region for Model S20 shown in Fig.\ref{fCurf2S20}, here at 
time 0.15 $\, t_{ff}$, when the peak density is $\log_{10}\left( \rho_{\rm max}/\rho_0\right)=9.3$, and the region 
is delimited by $(-0.6 \, \times R_0, 0.6 \, \times R_0)$ in the $x$-axis and by $(-0.7 \, \times R_0, -0.2 \, \times R_0)$ in 
the $y$-axis.}
\end{center}
\end{figure}

\begin{figure}
\begin{center}
\includegraphics[width=4.0 in]{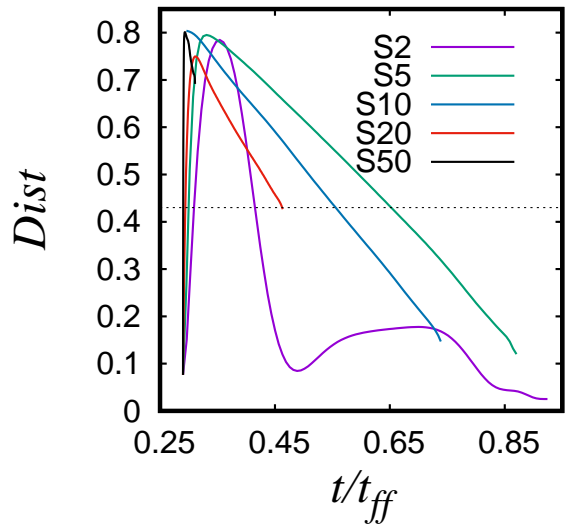}
\caption{\label{fDisCte} On the vertical axis, the distance from the shell of dense gas to the clump centre is shown, given in terms 
of the radius of the clump. The horizontal dashed line indicates the distance for models included in the iso-density 
plot shown in Fig.\ref{fMos4}.}
\end{center}
\end{figure}

\begin{figure}
\begin{center}
\includegraphics[width=4.5 in,height=6 in]{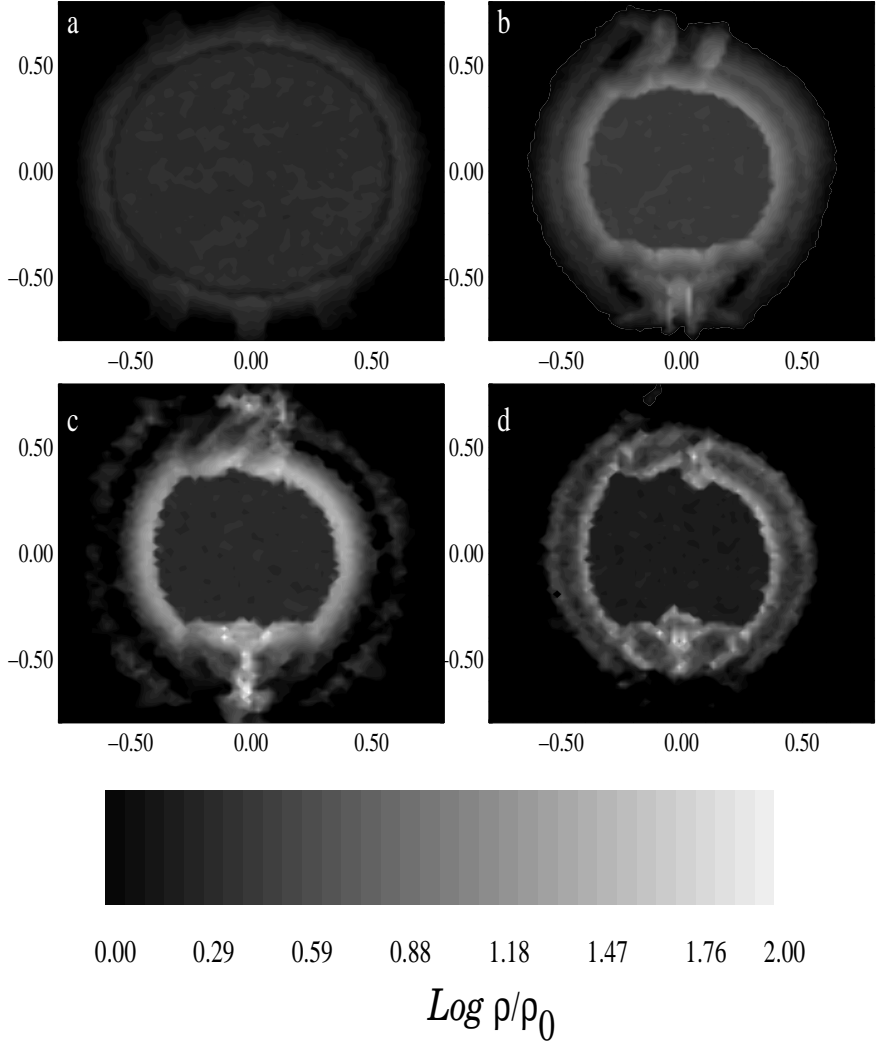}
\caption{\label{fMos4} Iso-density plot of the shock models S2-S20 for a region delimited by 
$(-0.8 \, \times R_0, 0.8 \, \times R_0)$ in the $x$-axis and by $(-0.8 \, \times R_0, 0.8 \, \times R_0)$ in the $y$-axis, when 
the particle distribution reaches a peak density of 
(a) $\rho_{\rm max}= 3.6 \times 10^{-20}\,$ g cm$^{-3}$ at time $t/t_{ff}=0.58$ for model S2; 
(b) $\rho_{\rm max}= 2.5 \times 10^{-19}\,$ g cm$^{-3}$ at time $t/t_{ff}=0.65$ for model S5; 
(c) $\rho_{\rm max}= 3.5 \times 10^{-17}\,$ g cm$^{-3}$ at time $t/t_{ff}=0.56$ for model S10 
and 
(d) $\rho_{\rm max}= 1.2 \times 10^{-10}\,$ g cm$^{-3}$ at time $t/t_{ff}=0.46$ for model S20. These 
snapshots are chosen to ilustrate the models shown in Fig.\ref{fDisCte} with a horizontal dashed line.}
\end{center}
\end{figure}

A last point to be emphasized is about the possibility of having fragmentation in the shock models studied 
in this paper. It seems that all the models have a certain tendency for the shell to fragment, as 
many cloudlets of dense gas are being formed along the shell. It is interesting to mention that there 
is another way in which fragmentation may occur. As was observed in Models S5, S10 and S20, some local 
turbulence develops in the outflow of gas in the negative direction of the $y$-axis. In Model S20, this 
effect is more noticeable, as can be seen in Fig.~\ref{fCurp2pS20Zi}, where a zoom-in of that 
region is shown.   

%%%%%%%%%%%%%%%%%%%%%%%%%%%%%%%%%%%%%%%

\section{Concluding Remarks}
\label{sec:conclu}

Three-dimensional, high-resolution, hydrodynamical simulations of a compressive velocity 
pulse impacting inward radially on a turbulent clump have been 
presented. The structure of the velocity pulse depends on several 
parameters, and the most important one, the magnitude of the initial velocity $V_0$,
has been varied systematically within the range 0--50 times the speed of sound 
$c_0$. This set-up was chosen to (i) reconsider and improve  
the one-dimensional radial simulations of \cite{gomez2007} to explore its 
feasibility as a mechanism of core formation; (ii) implement the idea of 
triggered collapse worked out by \cite{boss95} and \cite{vanhala} and more recently by \cite{henne}.    

All the shock models showed a separation into an external region $r> 0.95 \, R_0$ expanding outwards and an inner 
region falling inwards. This is a direct consequence of the form of the initial velocity pulse, which also has 
positive velocities for $r> 0.95 \, R_0$. The central regions are not affected by what 
is happening outside since the particles inside have no knowledge of the radially converging shock 
wave. Consequently, the 
central regions all proceed to collapse like the model without the velocity pulse for the duration of the 
simulation. Since each model runs for a different length of time, then the model with the highest 
Mach number (the shortest simulation time) had lower central densities since the collapse of the central 
regions has not gone on for so long.                  

Some of the main conclusions to be drawn from the shock models calculated here are the 
following: 

\begin{enumerate}

\item The turbulent clump collapses spontaneously a little before its free-fall time $t_{ff}$, so 
that its central region shows a lot of fragmentation, as is well known.

\item The velocity pulse speeds up the collapse of the turbulent clump, so that its new collapse times 
are shortened, as follows 0.31, 0.49, 0.69, 0.81 and  0.89 $\times \, t_{ff}$, for the models the models S50,S20,S10, S5 
and S2, respectively.

\item The pre-impact state of the clump in its central region is very similar to the outcome 
of the simulation with the lowest velocity pulse, Model S2, for which $V_0=2\, c_0$. The other 
shock models, Models S5--S50, exhibit a 
noticeable effect in changing the pre-impact configuration, so that a shell 
of dense particles is formed at the outermost region of the clump.

\item As expected, the physical properties of the shell, such as its position, width, mass, and dynamics, depend 
on the the magnitude of the initial velocity $V_0$.      

\item The mass contained in the shell is found to be large in general, ranging from 20--60 $M_{\odot}$, and
the higher the velocity $V_0$, the less mass is contained in the shell.

\item The central density of the shock models decreases as the velocity $V_0$ increases.

\item There is a dense gas structure in the central region of the shocked clump, which can be identified as 
a core, in the sense of \cite{gomez2007}, as its radial profile of density and velocity seem to indicate for Model 
S2. 

\item In agreement with the observation of \cite{vazquez2008}, in this paper there is evidence for 
the occurrence of a shock bouncing at the center only for model S2, even though it is likely that 
the other shock models will also produce a shock bouncing too, provided that that the evolution time 
be long enough.  

\item Additionally, there is a critical velocity of the pulse, such that for 
shock models with a lower pulse velocity (around 10 times the speed of sound) this 
dense central region will also form.       

\item The shock models studied here show ample possibilities of fragmentation, which can 
occur (i) along the shell of dense gas in the large velocity models, such as Model S50; (ii) in the small region of the outflow of 
gas, more noticeable in the moderate velocity models, such as Model S20; (iii) in the innermost central region of the shocked 
clump, where a cluster of small gas over-densities is formed only for the smallest velocity 
pulse, such as Model S2; so that this region is 
clearly inherited by the turbulent clump considered.      

\end{enumerate}   
%%%%%%%%%%%%%%%%%%%%%%%%%%%%%%%%%%%%%%%%%%%%%%%%%%%%%%%%%%%%%%%%%%%%%%%%%%%%
\acknowledgements

The author thankfully acknowledge the computer resources, technical 
expertise and support provided by the
Laboratorio Nacional de Superc\'omputo del Sureste de M\' exico through 
the grant number O-2016/047.
%%%%%%%%%%%%%%%%%%%%%%%%%%%%%%%%%%%%%%%%%%%%%%%%%%%%%%%%%%%%%%%%%%%%%%%%%%%%%%%%%%%%%%%%%%%%%%%%%%%%%%%%%%%%%%%% 
% Use this code if you wish to generate your bibliography with BibTeX;
% please replace first the string "an-demo" below with the name(s) of
% the BibTeX data base(s) you want to use.
% The resulting bibliography-output (the contents of the .bbl file)
% must be pasted into this file before submission.
% 
% \bibliographystyle{an}
% \bibliography{an-demo}
% 
% Replace the following example bibliography with your references
% before submission:

%\end{document}

%%%%%%%%%%%%%%%%%%%%%%%%%%%%%%%%%%%%%%%%%%%%%%%%%%%%%%%%%%%%%%%%%%%%%%%%%%%%%%%%%%%%%%%%%%%%%%%%%%%%%%%%%%%%
%\appendix

%\end{document}
    
%\clearpage
\newpage
%%%%%%%%%%%%%%%%%%%%%%%%%%%%%%%%%%%%%%%%%%%%%%%%%%%%%%%%%%%%%%

\begin{table}[ph]
\caption{The models, the values of their parameters and resulting configurations.}
{ \begin{tabular}{|c|c|c|c|c|c|c|c|}
\hline \hline
Model Number (label) & $V_0/c_0$ ($V_0$ km/s) & $\beta$   &  Figure                                & Configuration   \\
1 (C)                & 0.0 (0.0)         &       1.19     &  \ref{fCurf2final} and \ref{fCurf2}    & central cluster  \\
2 (S2)               & 2.0 (0.7)         &       0.1      &  \ref{fCurf2Sm2}                       & fragmented central region \\
3 (S5)               & 5.0 (1.7)         &       0.40     &  \ref{fCurf2S5}                        & a thick shell       \\
4 (S10)              &10.0 (3.4)         &       1.48     &  \ref{fCurf2S10}                       & a shell             \\
5 (S20)              &20.0 (6.8)         &       5.82     &  \ref{fCurf2S20}                       & a thin shell        \\
6 (S50)              &50.0 (17.1)        &       36.2     &  \ref{fCurf2S50}                       & a very thin shell   \\
\hline
\hline
\end{tabular} }
\label{tab:mod}
\end{table}
%%%%%%%%%%%%%%%%%%%%%%%%%%%%%%%%%%%%%%%%%%%%%%
%%%%%%%%%%%%%%%%%%%%%%%%%%%%%%%%%%%%%%%%%%%%
%%%%%%%%%%%%%%%%%%%%%%%%%%%%%%%%%%%%%%%%%%%%%
\newpage
%%%%%%%%%%%%%%%%%%%%%%%%%%%%%%%%%%%%%%%%%%%%%
%%%%%%%%%%%%%%%%%%%%%%%%%%%%%%%%%%%%%%%%%%%%
\newpage
%%%%%%%%%%%%%%%%%%%%%%%%%%%%%%%%%%%%%%%%%%%%%%%%%%%%
%%%%%%%%%%%%%%%%%%%%%%%%%%%%%%%%%%%%%%%%%%%%%%%%%%%%
\newpage
%%%%%%%%%%%%%%%%%%%%%%%%%%%%%%%%%%%%%%%%%%%%%%%%%%%%%%%%%%%%%%%
%%%%%%%%%%%%%%%%%%%%%%%%%%%%%%%%%%%%%%%%%%%%%%%%%%%%
\newpage
%%%%%%%%%%%%%%%%%%%%%%%%%%%%%%%%%%%%%%%%%%%%%%%%%%%%%%%%%%%%%%%
%%%%%%%%%%%%%%%%%%%%%%%%%%%%%%%%%%%%%%%%%%%%%%%%%%%%
\newpage
%%%%%%%%%%%%%%%%%%%%%%%%%%%%%%%%%%%%%%%%%%%%%%%%%%%%%%%%%%%%%%%
%%%%%%%%%%%%%%%%%%%%%%%%%%%%%%%%%%%%%%%%%%%%%%%%%%%%
\newpage
%%%%%%%%%%%%%%%%%%%%%%%%%%%%%%%%%%%%%%%%%%%%%%%%%%%%%%%%%%%%%%%
%%%%%%%%%%%%%%%%%%%%%%%%%%%%%%%%%%%%%%%%%%%%%%%%%%%%
\newpage
%%%%%%%%%%%%%%%%%%%%%%%%%%%%%%%%%%%%%%%%%%%%%%%%%%%%%%%%%%%%%%%
%%%%%%%%%%%%%%%%%%%%%%%%%%%%%%%%%%%%%%%%%%%%%%%%%%%%
\newpage
%%%%%%%%%%%%%%%%%%%%%%%%%%%%%%%%%%%%%%%%%%%%%%%%%%%%%%%%%%%%%%%
%%%%%%%%%%%%%%%%%%%%%%%%%%%%%%%%%%%%%%%%%%%%%%%%%%%%%%%%%%%%%%%
%%%%%%%%%%%%%%%%%%%%%%%%%%%%%%%%%%%%%%%%%%%%%%%%%%%%
\newpage
%%%%%%%%%%%%%%%%%%%%%%%%%%%%%%%%%%%%%%%%%%%%%%%%%%%
%%%%%%%%%%%%%%%%%%%%%%%%%%%%%%%%%%%%%%%%%%%%%%%%%%%%
%\end{document}
%%%%%%%%%%%%%%%%%%%%%%%%%%%%%%%%%%%%%%%%%%%%%%%%%%%%
\newpage
%%%%%%%%%%%%%%%%%%%%%%%%%%%%%%%%%%%%%%%%%%%%%%%%%%%%%%%%%%%%%%%
%%%%%%%%%%%%%%%%%%%%%%%%%%%%%%%%%%%%%%%%%%%%%
\newpage
%%%%%%%%%%%%%%%%%%%%%%%%%%%%%%%%%%%%%%%%%%%%%
%\end{document}
%%%%%%%%%%%%%%%%%%%%%%%%%%%%%%%%%%%%%%%%%%%%%
\newpage
%\end{document}
%%%%%%%%%%%%%%%%%%%%%%%%%%%%%%%%%%%%%%%%%%%%%
%%%%%%%%%%%%%%%%%%%%%%%%%%%%%%%%%%%%%%%%%%%%%%%%%%%%
%\end{document}
%%%%%%%%%%%%%%%%%%%%%%%%%%%%%%%%%%%%%%%%%%%%%%%%%%%%%
%%%%%%%%%%%%%%%%%%%%%%%%%%%%%%%%%%%%%%%%%%%%%%%%%%%%%%%%%%%%%%%%%%%%%%%%%%%%%%%%%%%%%%%%%%%%%%%%%%%%%%%%%%%%
\end{document}